\documentclass[prd,a4paper,aps,twocolumn,showpacs,
               nofootinbib,nobibnotes,superscriptaddress]
              {revtex4}

\usepackage{epsfig}
\usepackage{amsmath}
\usepackage{amssymb}

\newcommand{\ie}{{\em i.e. }}
\newcommand{\eg}{{\em e.g. }}
\newcommand{\cf}{\emph{cf.~}}

\newcommand{\HH}{{\cal H}}

\newcommand{\cd}{\cdot}

\newcommand{\lsim}{\lesssim}

\newcommand{\lag}{\langle}
\newcommand{\rag}{\rangle}

\newcommand{\al}{\alpha}

\newcommand{\de}{\delta}

\newcommand{\ga}{\gamma}

\newcommand{\La}{\Lambda}
\newcommand{\la}{\lambda}
\newcommand{\Om}{\Omega}
\newcommand{\om}{\omega}

\newcommand{\veps}{\varepsilon}

\newcommand{\be}{\begin{equation}}
\newcommand{\ee}{\end{equation}}
\newcommand{\bee}{\begin{equation*}}
\newcommand{\eee}{\end{equation*}}
\newcommand{\bea}{\begin{eqnarray}}
\newcommand{\eea}{\end{eqnarray}}
\newcommand{\bean}{\begin{eqnarray*}}
\newcommand{\eean}{\end{eqnarray*}}

\newcommand{\bk}{{\mathbf k}}
\newcommand{\bp}{{\mathbf p}}
\newcommand{\bq}{{\mathbf q}}
\newcommand{\bB}{{\mathbf B}}
\newcommand{\bx}{{\mathbf x}}
\newcommand{\by}{{\mathbf y}}
\newcommand{\bv}{{\mathbf v}}

\newcommand{\Gauss}{{\,\,{\rm Gauss}}}

\newcommand{\GeV}{\,\,{\rm GeV}}

\newcommand{\Mpc}{\,\,{\rm Mpc}}

\newcommand{\mr}[1]{\mathrm{#1}}
\newcommand{\rad}{{\rm rad}}

\newcommand{\TT}{\Delta\eta_*}
\begin{document}

\title{Gravitational waves from stochastic relativistic sources: \\
primordial turbulence and magnetic fields}

\author{Chiara Caprini}
\email{chiara.caprini@physics.unige.ch}
\affiliation{D\'epartement de Physique Th\'eorique, Universit\'e de
  Gen\`eve, 24 quai Ernest Ansermet, CH--1211 Gen\`eve 4, Switzerland}

\author{Ruth Durrer}
\email{ruth.durrer@physics.unige.ch}
\affiliation{D\'epartement de Physique Th\'eorique, Universit\'e de
  Gen\`eve, 24 quai Ernest Ansermet, CH--1211 Gen\`eve 4, Switzerland}

\date{\today}

\begin{abstract}

The power spectrum of a homogeneous and isotropic stochastic variable,
characterised by a finite correlation length, does in general not vanish on
scales larger than the correlation scale. If the variable is a
divergence free vector field, we demonstrate that its power spectrum is blue on
large scales. Accounting for this fact, we compute the gravitational waves
induced by an incompressible turbulent
fluid and by a causal magnetic field present in the early universe. The
gravitational wave power spectra show  common features: they are both blue on
large scales, and peak at the correlation scale. However,  
the magnetic field can be treated as a coherent source and it is active for a
long time. This results in a very  effective conversion of magnetic
energy in gravitational wave energy at horizon crossing.
Turbulence instead acts as a source for gravitational waves over a
time interval much shorter than a Hubble time, and the conversion into 
gravitational wave energy is much less effective. We also derive a strong
constraint on the amplitude of a primordial magnetic field when the
correlation length is much smaller than the horizon. 

\end{abstract}


\maketitle

\section{Introduction}

Gravitational waves, once emitted, propagate freely in spacetime
(affected only by the expansion of the universe); they do not
interact and are not absorbed. They hence provide a direct probe
of physical processes that took place at the time of their 
generation, in the very early universe. The potential detection of
a gravitational wave background of primordial origin provides an
observational signal of sources that have disappeared by now.
Primordial sources of gravitational waves are for example
inflation, cosmic topological defects, a pre big bang phase of
expansion,  primordial phase transitions, a phase of turbulence in
the cosmic fluid and primordial magnetic fields. Gravitational
waves produced during inflation~\cite{inf}, by topological
defects~\cite{top} or during a pre-big bang phase~\cite{pre} have
been studied extensively. In this paper we will consider the last
two sources which can also be associated with a phase transition.

Gravitational waves are most probably the only remnant from a
period of turbulence in the primordial plasma. In order to be maintained, 
turbulence necessitates a continuous injection of energy  in
the absence of which it dissipates rapidly leaving no other
trace behind. A possible source for this energy input could
come from the false vacuum  decay during a first order phase transition. 

Primordial magnetic fields
could be at the origin of the ubiquitous magnetic fields observed in
galaxies and clusters today. Their imprint in the form of
gravitational waves can confirm their presence in the primordial universe 
also on scales on which magnetic fields are completely dissipated today.

In the following we study in detail the generation of gravitational
waves by these two sources and we compare the results obtained.
The sources share common properties: they are both related to
divergence-free vector fields, and they both obey a relativistic
equation of state so that they evolve  { like radiation} with the
expansion of the universe. They are also both correlated over a finite
correlation length, if one accounts only for causally created
magnetic fields and turbulence. However, turbulence lasts for a
short time (\eg the duration of the phase transition) compared to the
evolution of the universe and it acts incoherently as a source of
gravitational waves, while the magnetic
field is a coherent source active in principle 
up to the time of matter and radiation equality.

In the next section we demonstrate that the power spectra of both
sources are blue on scales larger than the correlation scale. This
implies that turbulence in the primordial universe has a Batchelor
spectrum on large scales. Moreover, it gives rise to a background
of gravitational waves also on these large scales, which has been
disregarded so far. In section \ref{turbulence} we model the power
spectrum of turbulence and of the turbulent anisotropic
stress, and we analyse the creation of gravitational waves. We
propose a new method to perform this analysis, avoiding an error
in the gravitational wave dispersion relation which is found in
previous works. We find that the gravitational wave spectrum peaks at a
frequency corresponding to the \emph{size} of the largest turbulent eddies. 
In Section~\ref{magnetic} we compute the gravitational waves
induced by a causal magnetic field correlated on a scale which may
be much smaller than the horizon, and we constrain the magnetic fields using 
the nucleosynthesis bound. Finally, in Section~\ref{comparison}, we
compare the two results, 
concluding that a radiation-like, long acting, coherent source
like the magnetic field is more efficient in generating
gravitational waves on large scales than a rapidly dissipating,
incoherent source such as turbulence. We believe that this result
is valid beyond the two examples discussed here.

We consider a Friedmann universe with flat spatial sections,
\be
ds^2 =a^2(\eta)[-d\eta^2 +\de_{ij}dx^idx^j]\,.
\ee The variables ${\bf x }$, $r$, $L$ etc. denote comoving
distances and ${\bf k }$ is the comoving wave vector. The density
parameter is always scaled to today, $\Om_X (\eta)\equiv
\rho_X(\eta)/\rho_c(\eta_0)$, where the index $_0$ indicates the
present time. For relativistic species we have therefore
$\Om_X(\eta)=\Om_X(\eta_0)/a^4(\eta)$; we normalise $a(\eta_0)=1$ and
sometimes denote the present value of a density parameter simply by
$\Om_X(\eta_0) \equiv \Om_X$; likewise, $\rho_c=\rho_c(\eta_0)$. 
$\mathcal{H}=\frac{\dot{a}}{a}$ denotes the conformal Hubble parameter.  
The radiation energy density today is taken to be $\Om_{\rm
rad}(\eta_0)\equiv \Om_\rad =4.2 \times 10^{-5}$~\cite{WMAP}.
 We use the expression
$$ a(\eta) = H_0\eta\left(\frac{H_0\eta}{4} + \sqrt{\Om_{\rm rad}}\right) $$
for the scale factor which is strictly true only in a flat matter--radiation
universe. But since we are mainly concerned about the physics during
the radiation dominated phase of the universe, neglecting the
cosmological constant 
$\Lambda$ is irrelevant for our purpose. Values of conformal time
$\eta\ll 2(\sqrt{2}-1)\sqrt{\Om_{\rm rad}}/H_0=\eta_{\rm eq}$ denote
the radiation dominated phase, while during the matter dominated phase $\eta\gg
\eta_{\rm eq}$. 

\section{The power spectrum of divergence free vector fields at large scales}
\label{powerspectrum}
We want to discuss gravitational waves generated from the energy
momentum tensor of divergence-free vector fields. In this section we
derive the generic form of the spectrum for such vector fields.

We consider a stochastic vector field $\bv(\bx)$, statistically homogeneous and
isotropic, hence $\langle v_i(\bx)\rangle =0$. The general form of its
two point correlation tensor is (see for
example \cite{LL})~\footnote{
This form of the correlator is not completely general in the
sense that it does not contain helicity, i.e. a component which is odd
under parity (see~\cite{Tina}). We neglect parity violating terms in
this work.}
\bea
b_{ij}(\bx)&\equiv&\lag v_i(\by) v_j(\by+\bx)\rag= \nonumber \\
&=&\Sigma(r)(\de_{ij}-\hat{x}_i\hat{x}_j)+\Gamma(r)\hat{x}_i\hat{x}_j\,,
\label{bijr}
\eea
where $\hat{\bx}$ is the unit vector in direction $\bx$, and $r$ is the
length of $\bx$. Here $\Sigma(r)$ and $\Gamma(r)$ denote the correlation functions in
directions perpendicular and parallel to $\hat{\bx}$. If the
stochastic vector field is divergence-free (for example in the case of the
turbulent velocity field of an incompressible fluid, or in the case of a magnetic field), 
the correlation functions $\Sigma(r)$ and
$\Gamma(r)$ are related. It
is easy to verify that $\frac{\partial v_i(\bx)}{\partial x_i}=0$ yields
\be
\frac{d\Gamma(r)}{dr}=\frac{2}{r}\left(\Sigma(r)-\Gamma(r)\right)\,.
\label{nulldivergence}
\ee
Homogeneity implies that the Fourier transform of $\bv$ for different
wave vectors $\bk\neq \bq$ is uncorrelated, so that the power spectrum
is of the form
\bea
&&\lag v_{i}(\bk)v_j^*(\bq)\rag = (2\pi)^3 \de(\bk-\bq)P_{ij}(\bk)\,,\\
&& \mbox{where} \qquad P_{ij}(\bk)=\int d^3 x \, e^{i\bk\cdot\bx}b_{ij}(\bx)\,.
\label{powv}
\eea
Isotropy and $\mathbf{\nabla}\cd{\bf v} = 0$ require the following
simple form for the power spectrum $P_{ij}(\bk)$
\be
P_{ij}(\bk)=P(k)(\de_{ij}-\hat{k}_i\hat{k}_j)\,.
\label{vpowerspectrum}
\ee
The trace of Eq.~\eqref{powv} together with Eq.~\eqref{bijr} lead to
\be
 P(k)=\int d^3 x \, e^{i\bk\cdot\bx} \left(\Sigma(r) +
 \frac{1}{2}\Gamma(r) \right)~. \label{PSiGa}
\ee
In the expanding (non inflationary) universe, any causal random process is characterised
by a finite correlation length $L$, which is at most the size of the
horizon at the epoch at
which the process is occurring, $L\leq\eta_*$. The correlation function
$b_{ij}(\bx)$ is generally expected to decay at large distances, but in
the case of the expanding universe, causality  even requires
that it vanishes identically on scales larger than the horizon $\eta_*$
(if $v_i(\bx)$ is a classical field, see \cite{gabrielli}).
Therefore, in the case of interest to us, the correlation function
$b_{ij}(\bx)$ of a causal process is always a function with compact
support. This implies that its Fourier transform must be an analytic function
in the variable $\bk$. Hence, the power spectrum (\ref{vpowerspectrum})
must be analytic in $\bk$. In order for
$P(k)(\de_{ij}-\hat{k}_i\hat{k}_j)$ to be an analytic function of the
components $k_i$ also at $\bk=0$,  we have to impose
$P(k\rightarrow 0)\propto k^n$, with $n$ an even positive integer, $n\geq 2$.
From Eqs. (\ref{powv}) and
(\ref{vpowerspectrum}) it follows that $P(k)$ is given by
\be
P(k)=\frac{1}{2}\int d^3 x\, e^{i\bk\cdot\bx}b_{ii}(r)=2\pi\int_0^\infty dr
\,r^2\, \frac{\sin(kr)}{kr}\,b_{ii}(r)\,.
\label{Pkdef}
\ee
For small $k$ we obtain up to order $k^2$
\bea
P(k\rightarrow 0)&=&2\pi\int_0^\infty dr\, r^2 \,b_{ii}(r)\nonumber\\
&-&\frac{\pi}{3}\,k^2\int_0^\infty dr\, r^4\, b_{ii}(r)+\mathcal{O}(k^4)\,.
\label{Pk0}
\eea
Analyticity of $P_{ij}(\bk)$ requires the first term on
the right hand side of Eq.~(\ref{Pk0}) to vanish. This result 
can also be obtained
directly from Eq.~(\ref{nulldivergence}) using the fact that $
  \Gamma(r)$ vanishes for sufficiently large distances $r$.
Indeed, using (\ref{nulldivergence}) one finds
\be
\int_0^\infty dr\, r^2\,b_{ii}(r)=r^3 \Gamma(r)|^{\infty}_0 = 0\,,
\ee
since $\Gamma(r)$ has compact support. One is thus left with
\be
P(k\rightarrow 0)=\left(\frac{2\pi}{3}\int_0^\infty dr\,
r^4\,\Gamma(r)\right)\, k^2 +\mathcal{O}(k^4) \,.
\label{turbulentspectruminzero}
\ee

These considerations apply both to the power spectrum of a magnetic field,
if it has been generated in the early universe by a causal process
\cite{magcausalgeneration}, and to the power spectrum of turbulent
motions in the primordial fluid, which can be generated for example at
the end of a first order phase transition (if one assumes that the
fluid is incompressible) \cite{kamionkowski,kosowsky,dolgov,nicolis}.

\section{Gravitational waves from turbulence}
\label{turbulence}

A period of turbulence in the primordial universe can leave an observable
trace today in the form of a stochastic background of gravitational waves.
The most natural source of turbulence in the early
universe is a first order phase transition (see \cite{dolgovgrasso} for
another potential source of turbulence), where bubbles of the broken
phase nucleate and expand into the symmetric phase. Therefore
we concentrate on this case. We  parameterise the specifics of the
phase transition following~\cite{kamionkowski,kosowsky,dolgov,nicolis}.
The temperature and the value
of the scale factor at which it takes
place are denoted by $T_*,~ a_*$; the comoving size of the largest bubbles at
the moment when they collide is $L$. This length corresponds to
the largest scale on which turbulence 
develops (the size of the largest eddies).
The ratio of the false vacuum energy density to the thermal energy density of
the universe at the moment of the transition is $\alpha=\rho_{\rm
vac}/\rho_*$;  and the velocity of the bubble walls $v_b(\alpha)$.
Finally, the fraction of vacuum energy density which is converted into
kinetic energy, turbulence, of the fluid is denoted by $\kappa=\rho_{\rm
  kin}/\rho_{\rm vac}$.

To generate a considerable amount of gravitational waves, one needs to move 
big masses rapidly. This can be achieved when a first order
phase transition proceeds through detonation of the broken phase
bubbles. In this case the bubble walls propagate faster than the speed
of sound  $1/\sqrt{3}<v_b(\alpha)<1$, and there is a large
concentration of energy at the bubble walls. The phase transition is
completed in a (conformal) time interval of the order of
$\TT \simeq L/v_b(\alpha)$ which is much smaller than one Hubble time
$\TT \mathcal{H}_*\ll 1$. Typical values for $\TT \mathcal{H}_*$ for the
electroweak phase transition are $0.001-0.01$
(see~\cite{kamionkowski,kosowsky,dolgov,nicolis} and references therein).

A first order phase transition causes
injection of kinetic energy in the primordial fluid at a
characteristic scale $L$ corresponding to the size of the largest bubbles
at the moment when they collide. Therefore, the primordial plasma
is stirred on the scale $L\ll\mathcal{H}_*^{-1}$.
Since the Reynolds number is very high in the
early universe, turbulent motions are generated \cite{viscosity}.
We assume that the generated turbulence is homogeneous, isotropic, fully
developed and stationary. Then, since $L$ is the principal turbulence
scale, on smaller scales (larger wave numbers) $k> L^{-1}$ an energy
cascade is established
which proceeds from large to small scales. Eventually one expects the
formation of a Kolmogorov spectrum in the `inertial range'
$L^{-1}\ll k \ll \la^{-1}$, where $\la$ denotes the dissipation
scale~\cite{LL}.  In the next section we derive an
approximate form of the turbulent power spectrum which is valid at all
wavenumbers (including large scales, small wavenumbers $k<L^{-1}$).

\subsection{Turbulence power spectrum}

The variable $v_i(\bx)$ represents the fluid velocity at the position $\bx$,
and we denote its power spectrum by $P_v(k)$.
Following \cite{kosowsky} we define the total kinetic energy per unit
enthalpy of the fluid as
\be
\frac{\rho_{\rm kin}}{\rho_*+p_*}\equiv
\frac{1}{2}\lag v_i(\bx) v_i(\bx)\rag = \frac{1}{2}b_{ii}(0)\,,
\label{kinenergy}
\ee
where $\rho_*$ and $p_*$ denote the energy density and pressure at the time
of the turbulence. This non-relativistic approximation is strictly valid only 
for $\lag v^2\rag\ll 1$: the Kolmogorov theory of turbulence has been
formulated and verified only in this regime. However, 
we shall use it also in the relativistic regime, imposing the upper bound 
$\lag v^2\rag \leq 1/3$ (\cite{steinhardt}) 
where it still gives the correct order
of magnitude. The details of the
Kolmogorov spectral shape are probably not valid in the relativistic
regime, but we follow here earlier literature on the
subject, see~\cite{kamionkowski,kosowsky,dolgov,nicolis}. 
As usual, we also define the energy dissipated per
unit time and unit enthalpy by~\cite{LL,hinze,davidson}
\be
\veps=-\frac{d}{d\eta}\frac{\lag v^2 \rag}{2}\,.
\label{epsilon}
\ee

The turbulent velocity power spectrum can be evaluated from
Eq.~(\ref{Pkdef}), if the shape of the two point correlation
function $b_{ii}(r)$ is known.
If the turbulence is homogeneous and isotropic, in the
inertial range $\la\ll r\ll L$ the correlation function is given by the
Kolmogorov two thirds law $(\varepsilon r)^{2/3}$. In the dissipation range
$r\ll \la$, the fluid motions are regular and one can Taylor expand the
velocity field as a function of $r$. This entails that the correlation function
is proportional to $r^2$ \cite{LL}. Moreover, the symmetric tensor
field $b_{ij}(\bx)$
has compact support and its divergence vanishes. On this basis,
we make the following Ansatz for the transverse correlation function
\bea
\Sigma(r) \simeq \left\{ \begin{array}{ll}
\frac{\lag v^2\rag}{3}+\Sigma_0\big(\frac{r}{\la}\big)^2 &
\text{for}~0<r\leq \la\\
\frac{\lag v^2\rag}{3}+\Sigma_0\big(\frac{r}{\la}\big)^{2/3}
& \text{for}~\la\leq r< L\\
0 &\text{otherwise}\,,
\end{array} \right. \label{sigmar}
\eea
with $\Sigma_0$ an as yet unspecified constant. From the
divergence-free condition
(\ref{nulldivergence}) we now determine the radial correlation function
\bea
\Gamma(r) \simeq \left\{ \begin{array}{ll}
\frac{\lag v^2\rag}{3}+\frac{\Sigma_0}{2}\big(\frac{r}{\la}\big)^2 &
\text{for}~0<r\leq \la\\
\frac{\lag v^2\rag}{3}-\frac{\Sigma_0}{4}\big(\frac{\la}{r}\big)^2
+\frac{3}{4}\Sigma_0\big(\frac{r}{\la}\big)^{2/3} & \text{for}~\la\leq r \leq L\\
0 &\text{otherwise}\,.
\end{array} \right. \nonumber
\eea
Eq.~(\ref{nulldivergence}) also requires that $\Gamma$ be continuous,
so that $\Sigma$ does not contain a singularity ($\Gamma$ may, however, have
a kink which leads to a discontinuity in $\Sigma$). The continuity of
$\Gamma(r)$ fixes $\Sigma_0$ in terms of $\la$, $\lag v^2\rag$ and $L$:
$\Gamma(r\geq L)=0$ implies
\be
\Sigma_0\simeq -\frac{4}{9}\left(\frac{\la}{L}\right)^{2/3}\lag v^2\rag
\ee
(where we have neglected a term of order $(\la/L)^2$).
The radial correlation function
is non negative, continuous and identically zero on scales larger
than the correlation scale. The transverse correlation function starts
positive, passes through zero inside the inertial range at
$r=(3/4)^{3/2}L$, and has a jump of $-(1/9)\lag v^2\rag$ at $r=L$ in
order to satisfy the
condition of compact support. This jump corresponds to a kink in $\Gamma(r)$
for $r=L$. The integral of $b_{ii}(r)$ over space
vanishes, so that the power spectrum satisfies
$P_v(k=0)=0$. Fourier transforming the correlation function given
above, one finds
\bea
\vspace*{0.2cm}
& & P_v(k)\simeq 2\pi \lag v^2\rag \times \nonumber \\
& &\left\{ \begin{array}{ll}
\vspace*{0.2cm}
\frac{2}{765} L^5 k^2 &\text{for}~k\ll L^{-1}\\
\vspace*{0.2cm}
\frac{55}{81}\sqrt{3} \Gamma\big(\frac{2}{3}\big)
L^{-2/3}k^{-11/3} &\text{for}~L^{-1}\ll k\ll \la^{-1}\\
0 &\text{otherwise}\,.
\end{array} \right. \label{velocityspectrum}
\eea
In the inertial range we recover the Kolmogorov spectrum,
but the power is non-zero also for small wave-numbers $k<L^{-1}$ (a
well known result in the theory of turbulence \cite{hinze,davidson},
but which has not been taken into account
in cosmological applications \cite{kosowsky,dolgov}\footnote{In
  Refs.~\cite{shiro,son,hind,campa,axel,arjum} the spectrum at large
  scales is given, but 
there  MHD turbulence, not pure fluid turbulence is considered. Furthermore,
the fact that for a  divergence free vector field causality implies
that the spectrum is blue, $P_v\propto k^2$ for $Lk\ll 1$ is
only acknowledged in Ref.~\cite{hind}. If we actually follow the
renormalization group analysis outlined in \cite{arjum}, we find that
at large scales, $kL\ll 1$, the driving force has a $k^4$ spectrum and
the stable fixed point has the critical exponents $z=2$ and $\chi
=-3/2$. This can be inferred simply from causality and from the fact that
the field is divergence free.} so far).

The discontinuity of the spectrum at the peak $k=L^{-1}$ is not
real. It is a consequence of our approximations which are valid only
for $kL\ll 1$ and $kL\gg 1$ respectively. A numerical calculation of the
spectrum would smoothly turn from the $k^2$ behavior for $kL\ll1$ to the
$k^{-11/3}$ for $kL\gg 1$. The discontinuity indicates, however, that the peak
amplitude obtained by extrapolating the asymptotic behavior right up
to $k=1/L$ is not very reliable. Eq.~(\ref{kinenergy}) gives the spectrum
normalisation 
\be
\frac{\lag v^2\rag}{2}=\frac{b_{ii}(0)}{2}=\int \frac{d^3k}{(2\pi)^3}P_v(k)\,.
\ee
Integrating Eq.~(\ref{velocityspectrum}), we obtain instead $0.76 \lag
v^2\rag$. This overestimation of the total energy is also  
due the extension of our approximations up to the peak
$k=L^{-1}$. It indicates that the peak amplitude might be roughly a
factor of $2$ too high. In the following, we will neglect these
discrepancies which are within the accuracy of our approximations.

We define the ratio of the total kinetic energy 
of turbulence to the radiation energy density 
(\cf Eq.~(\ref{kinenergy})): 
\bea
\frac{\Om_T(\eta_*)}{\Om_\rad(\eta_*)}  
&\equiv& \frac{1}{\Om_\rad(\eta_*)}\int_0^\infty \frac{dk}{k}
\frac{d\Om_T(k,\eta_*)}{d\log(k)}\nonumber \\
&=&
 \frac{\rho_\mathrm{kin}}{\rho_*}=\frac{2\lag v^2\rag}{3}\simeq
\int \frac{d^3 k}{6\pi^3}P_{v}(k) \, .
\label{vsquared}
\eea
The above equation defines the turbulent energy spectrum per logarithmic unit
of frequency
\be
\frac{d\Om_T(k,\eta_*)}{d\log(k)}\simeq\Om_\rad(\eta_*)\frac{2}{3\pi^2}k^3 P_v(k)\,.
\label{turbulentenergyspectrum}
\ee
Usually the turbulent energy spectrum is defined as $E(k)=k^2P_v(k)/(2\pi^2)$
\cite{LL}, in our notations we have 
\be
\frac{d\Om_T(k,\eta_*)}{d\log(k)}=\Om_\rad(\eta_*)\frac{4}{3}kE(k)\,.
\ee
In the Kolmogorov inertial range the energy spectrum has the form
$E(k)=C\varepsilon^{2/3}k^{-5/3}$ \cite{LL}, following \cite{kosowsky} 
we set the value $C=1$. Comparing $E(k)$ and the power spectrum 
(\ref{velocityspectrum}) we find the relation
\be\label{eq19}
\lag v^2 \rag\simeq 2(\varepsilon L)^{2/3}\,.
\ee

On large scales $k< L^{-1}$ the above considerations on
the properties of null divergence and compact
support of the correlation function imply the formation of a Batchelor
spectrum for the turbulent energy.
The definition of the energy spectrum (\ref{turbulentenergyspectrum})
and Eq.~(\ref{turbulentspectruminzero}) give
\be
\frac{d\Om_T}{d\log(k)}(k\rightarrow 0) \simeq \Om_{\rad}(\eta_*)\,\frac{4}{9\pi}\,I\,k^5 \,,
\ee
where $I=\int_0^\infty dr\,r^4\,\Gamma(r)$ is Loitsyansky's integral
\cite{LL,hinze,davidson}.
Since the correlation function has  compact support,
the Navier-Stokes equation requires that $I$ be constant in time.
Therefore, the turbulence
is persistent on large scales and its rate of decay increases with wavenumber.
Loitsyansky's integral is a measure of the angular
momentum of the turbulent fluid. Its constancy is a manifestation of the
conservation of angular momentum~\cite{LL}. A simple estimate gives
$I\sim \lag v^2\rag L^5$. Exploiting the constancy of $I$ and the energy decay
equation $\frac{d\lag v^2\rag}{d\eta}\sim -\frac{\lag v^3\rag}{L}$, 
Kolmogorov has estimated the decay of the kinetic energy in freely evolving
turbulence $\lag v^2\rag(\eta)\propto \eta^{-10/7}$ and the growth of the correlation
length $L(\eta)\propto \eta^{2/7}$~\cite{LL,hinze,davidson}.
In some cases these decay laws are a reasonable fit to the experimental data,
nevertheless nothing tells us that they are verified for our
oversimplified model of turbulence in the primordial fluid.
In the following, we will simply neglect the overall decay of
turbulence as far as gravitational wave production is concerned.
We shall assume that turbulence appears at some time $\eta_*$, remains
active for a short period $\tau$ and simply disappears
afterwards. We shall not model its decay in any detail.

\subsection{Eddy turnover and the time of duration of turbulence}

The energy cascade involves a hierarchy of vortices of different
sizes. To each of these structures on a given scale one can associate
a characteristic velocity, the eddy velocity. 
Following Refs.~\cite{kosowsky,dolgov}, we define the
characteristic eddy velocity at the largest scale $L$ by
\bea
v_L^2 &=& \frac{3}{2}\frac{1}{\Om_\rad(\eta_*)}\int_{L^{-1}}^{\la^{-1}}
  \frac{dk}{k} \frac{d\Om_T(k,\eta_*)}{d\log(k)} \nonumber \\ 
&\simeq& \lag v^2\rag  =
  \frac{3}{2}\frac{\Om_T(\eta_*)}{ \Om_\rad(\eta_*)} \,,
\label{vL}
\eea
where the second equality holds since most of the turbulent energy resides
in the large wavelengths $k\simeq L^{-1}$. The velocity of the largest
eddies, $v_L$, 
is another parameter that measures the energy stored in turbulence. 
For the maximal eddy velocity, $v_L^2 = 1/3$ we have
$\frac{\Om_T(\eta_*)}{ \Om_\rad(\eta_*)} = \frac{2}{9}$. In this
regime, our non-relativistic approximation underestimates the kinetic
energy of turbulence somewhat.

The velocity of the eddies on a length scale $\la<\ell<L$ is simply
\be
v_\ell=v_L\,\left(\frac{\ell}{L}\right)^{1/3}\,.
\ee
One further defines the eddy turnover time $\tau_\ell=\ell/v_\ell$ and
the turnover frequency $\om_\ell=1/\tau_\ell$.

This picture is valid only on scales $\la<\ell< L$. The principal scale
of turbulence, $L$, defines the extent of the region over which
velocities are correlated, and it is interpreted as the maximal size
of turbulent eddies.
On scales larger than $L$ there is no correlated fluid motion and therefore no
eddies. Similarly on scales smaller than $\la$ the Reynolds number is of
order unity and the fluid is not in a turbulent regime, therefore no eddies are
formed, and we do not associate a
turnover frequency on scales larger than $L$ or smaller than $\la$.
However, the energy distribution is not zero for $\ell<\la$
and $\ell>L$ since the eddies of a given size $\ell$ contribute to the
power spectrum $P_v(k)$ across the full range of wavenumbers\footnote{In the
previous treatment (\cf Eq.~(\ref{velocityspectrum})), we have neglected
power on scales smaller than $\la$.
The dissipation scale contains very little energy, and is too small to be
relevant in the context of a cosmological treatment.}~\cite{davidson}.

The eddy turnover time on a given scale defines the time interval over
which the eddies of
 that size break down into smaller ones, transferring energy from
the original scale to a smaller one. This is the principle of the direct
cascade. If the stirring of the fluid is continuous, fully developed
and stationary
turbulence is established in a time interval of the order of the turnover
time on the largest scale $\tau_L$. Once the stirring stops, the turbulence
decays away in the same time interval. If the stirring lasts for less
than $\tau_L$, the
cascade of energy develops but the turbulence is not stationary. In Ref.
\cite{kosowsky} it is argued that, from the point of view of
gravitational wave production, this case can be modelled like
fully developed stationary turbulence lasting for a time $\tau_L$
(instead of for the stirring time). The assumption which goes
into this model is that eddies of every size $\ell<L$ are
generated and dissipated in the time interval  $\tau_L$ (so that they make
several oscillations before transferring energy to a smaller scale). Therefore,
we follow \cite{kosowsky} and define the total time of duration of
turbulence
as $\tau={\rm max}(\tau_L,\TT)$. We take here $\TT$ because
the external stirring lasts until the phase transition is completed.

For a first order phase transition this always reduces to
$\tau=\tau_L$ \cite{kosowsky}. From
Eqs.~(\ref{vL}) and (\ref{eq19}) we have
\be
v_L\simeq \sqrt{2}(\varepsilon L)^{1/3}\,,\hspace{1cm}
\tau_L \simeq \frac{1}{\sqrt{2}}\frac{L^{2/3}}{\varepsilon^{1/3}}\,.
\label{vLtauL}
\ee
Since $\rho_{\rm kin}=\kappa \alpha \rho_{*}$, $p_*=\rho_{*}/3$,
Eq. (\ref{epsilon}) gives
\be
\varepsilon \simeq \frac{3}{4}\frac{\kappa\alpha}{\tau}\,.
\ee
Let us assume that the stirring lasts for much longer than the
turnover time of the largest eddies. Then $\TT$ should be substituted
in the above equation for $\varepsilon$, so that
\be
\tau_L=\frac{1}{\sqrt{2}}\left( \frac{4}{3}\right)^{1/3}
\frac{L^{2/3}(\TT)^{1/3}}{(\kappa\al)^{1/3}}\,.
\ee
With this expression and using that $L\simeq
v_b(\alpha)\TT$,
the condition $\TT\gg \tau_L$ can be rewritten in terms of the
phase transition parameters,
\be
\TT\gg \tau_L\quad \Rightarrow \quad
\frac{v_b(\al)}{\sqrt{\alpha}}\ll \sqrt{\frac{3\kappa}{4}}\,.
\label{condition}
\ee
$v_b(\al)$ and $\kappa(\al)$ are given in Ref. \cite{kosowsky}:
\bea
v_b(\al)&=&\frac{1}{1+\al}\left(\frac{1}{\sqrt{3}}+\sqrt{\al^2+\frac{2}{3}\al}\right)\,,\\
\kappa(\al)&=&\frac{1}{1+0.72\al}\left(0.72\al+\frac{4}{27}\sqrt{\frac{3\al}{2}}\right)\,.
\eea
Substituting these expressions in (\ref{condition}) it turns out that
the condition $\TT\gg \tau_L$ is never verified for $\alpha\leq 1$.
Therefore, even in the case of a strongly first order phase transition,
the turbulence persists for a duration of the order of the
turnover time of the largest eddies. In the following analysis
we set $\tau\simeq\tau_L$.
During this time we neglect the decay of the turbulent kinetic energy and
the subsequent time dependence of the power spectrum $P_v(k)$. We assume that
the turbulence is stationary during the time interval $\tau_L$, and
instantaneously dissipated just afterwards.

\subsection{Anisotropic stress power spectrum}
\label{turbuanisostress}
We now evaluate the anisotropic stress generated by the turbulence.

The energy momentum tensor of the
turbulent fluid has an anisotropic stress component that induces
tensor perturbations $h_{ij}$ of the metric. The propagation equation
for tensor perturbations in a radiation dominated universe is
\be
\ddot{h}_{ij}({\mathbf k},\eta)+\frac{2}{\eta}
\dot{h}_{ij}({\mathbf k},\eta)+k^2h_{ij}({\mathbf k},\eta)
= 8\pi G a^2(\eta)\Pi_{ij}({\mathbf k},\eta)\label{gwevolution}\,,
\ee
where $\Pi_{ij}({\mathbf k},\eta)$ is the anisotropic stress.
Tensor perturbations are induced on each mode $\bk$. If
the mode is super-horizon at the moment of generation,
the last term on the left hand side of the above
equation is much smaller than the damping term $\frac{2}{\eta}
\dot{h}_{ij}$ and the tensor perturbation is not yet
oscillating. Once the perturbation enters the horizon, the oscillatory term
can no longer be neglected and the perturbation becomes a gravitational wave in
the proper sense. 

If the source of the anisotropic stress
has a blue power spectrum on large scales $k<L^{-1}$, then the
anisotropic stress power spectrum is white noise (and not zero) for every
$k<L^{-1}$ as we now show. At the present time the modes
$H_0<k<L^{-1}$ are sub-horizon. 
On these scales there is a background of gravitational waves
coming from the primordial source, characterised by a white noise
spectrum. In the case of turbulence after a phase transition, this signal has
not been calculated in previous analysis.

The energy momentum tensor of the fluid is
\be
T_{\mu\nu}=(\rho+p)\,v_\mu v_\nu+p \,g_{\mu\nu}\,.
\label{fluidtensor}
\ee
We extract the spatial, transverse and traceless part which
is relevant for tensor perturbations.
The turbulence is active on sub-horizon scales $k\gg 1/\eta_*$ and for a time
much shorter than one Hubble time. For this short period of time, the
expansion of the
universe can be neglected, and we may assume a Minkowski background.
 In Eq.~(\ref{fluidtensor}) only the second order term
$T_{ij}(\bx,\eta_*)= 4/3 \,\rho_*\, v_i(\bx,\eta_*)\,v_j(\bx,\eta_*)$ generates
tensor perturbations~\cite{kosowsky}.
We have substituted $p_*=\rho_*/3$
since the phase transition takes place
during the radiation dominated era. In $\bk$--space, the tensor
anisotropic stress is  given by
$\Pi_{ij}(\bk)=(P_{il}P_{jm}-1/2P_{ij}P_{lm})T_{lm}(\bk)$
where $P_{ij}=\delta_{ij}-\hat{k}_i\hat{k}_j$ is the transverse
traceless projector, and
$T_{ij}(\bk,\eta_*)=\frac{4}{3}\rho_*
\int\frac{d^3q}{(2\pi)^3}\,v_i(\bq,\eta_*)\,v_j(\bk-\bq,\eta_*)$, up
to an irrelevant trace which vanishes in the projection.
The anisotropic stress power spectrum
$\lag\Pi_{ij}(\bk,\eta_*)\Pi_{ij}^*(\bq,\eta_*)\rag$ has already been
calculated in Ref. \cite{kosowsky}: using Wick's theorem to
reduce the four point spectral function of the velocity, one obtains
\be
\lag\Pi_{ij}(\bk,\eta_*)\Pi^*_{ij}(\bq,\eta_*)\rag
=\left(\frac{4\rho_*}{3}\right)^2 \de(\bk-\bq) \Pi_v(k,\eta_*)\,,
\label{correlatorPi}
\ee
\be
\Pi_v(k,\eta_*)=\int d^3p P_v(p,\eta_*)P_v(|\bk-\bp|,\eta_*)
(1+\gamma^2)(1+\beta^2)\,,
\label{Pispectrumint}
\ee
with $\gamma=\hat{\bk}\cdot\hat{\bp}$, $\beta=\hat{\bk}\cdot\widehat{\bk-\bp}$.
In principle, we are not allowed to use Wick's theorem, the
turbulent velocity field being not at all Gaussian. We invoke it
 here for simplicity and to obtain an expression for the
four-point function. We hope that even if not correct in detail, this
gives the right order of magnitude for the four point
function. Reducing the four point spectral function to the power
spectrum by means of Wick's theorem
is sometimes used as a closure method also in more detailed studies
of turbulence~\cite{davidson}.

The anisotropic stress power spectrum is explicitly
calculated at the time of the phase transition $\eta_*$. The time evolution
equation for the turbulent power spectrum is highly non linear
and involves the three point function~\cite{LL}.
As already mentioned in Section~\ref{powerspectrum},
in our approximate approach to the evaluation of gravitational
waves generated from the turbulence, we ignore the detailed time
dependence of the anisotropic stress, since it is not very relevant
for  the resulting gravitational waves which are given by a time
integral of the source. This however does not mean that we are
treating the turbulence as a coherent source, as will become clear in the next
section.

The integral in (\ref{Pispectrumint}) cannot be done exactly, but good
approximations have already been derived in Ref.~\cite{kosowsky} for small
scales $k>L^{-1}$, and in Refs.~\cite{nucleosintesi,helicity} for large scales
$k<L^{-1}$. With Eqs. (\ref{velocityspectrum}) and (\ref{vsquared}), 
we find\footnote{
The analytical approximations used in the above
mentioned references are not entirely correct, because of the bounds that
have been used in the angular integration.
$\ga=\hat{\bk}\cdot\hat{\bp}$ is not varying from $-1$ to $1$ but its bounds
depend on $\bk$ and $\bp$. We have checked that using the correct
bounds does not significantly modify the approximated final result.}
\bea
\Pi_v(k)\simeq \frac{27}{8}\pi^4
\left(\frac{\Om_T(\eta_*)}{\Om_\rad(\eta_*)}\right)^2 \left\{ 
\begin{array}{l}
\frac{2}{13}L^{3}  \\  \text{for}~k < L^{-1} \vspace*{0.2cm} \\
L^{-2/3}\,k^{-11/3} \\
 \text{for}~L^{-1}< k< \la^{-1} \vspace*{0.2cm} \\
0 \quad \text{otherwise}\,.
\end{array} \right. \label{Pispectrum}
\eea
As anticipated,  the tensor source power spectrum is white noise on
large scales. However, because of the convolution, the largest
contribution to $\Pi_v(k)$ on these scales does not come from the velocity
power spectrum on the same scales, but from the value of the velocity power
spectrum at the maximum, $k\simeq L^{-1}$.
The discontinuity of the spectrum at $k=L^{-1}$ is not physical, it is due
to the approximations we used in evaluating integral~(\ref{Pispectrumint}).

\subsection{Generation of gravitational waves}
\label{GWturbulence}

We now determine the spectrum of
gravitational waves which are induced by the turbulent motion of
the primordial fluid. The turbulent velocity is a random variable,
but within Richardson's model of the energy cascade it is possible
to associate a frequency to the fluid motions at each scale, since the fluid
motions are vortical. This `oscillatory' behaviour  is
peculiar to each realisation of turbulence, and is
lost in the statistical average. The velocity power spectrum $P_v(k)$
is not oscillating in time. First of all, the Fourier transform decomposes the
velocity field into a hierarchy of waves, not of eddies. Moreover, a
statistical model cannot reproduce the detailed dynamics of individual vortex
events. However, the `oscillatory' behaviour of the source is relevant
for the generation of the gravitational wave background. 

In order to account for these oscillations  which are
present in turbulent motion on scales smaller than $L$,
we model the random velocity of the turbulent fluid heuristically as
\bea
v_i(\bk,\eta)=\left\{ \begin{array}{ll}
v_i(\bk) & {\rm for}~k<L^{-1}\\
v_i(\bk)e^{i\omega_k \eta} & {\rm for}~L^{-1}<k<\la^{-1}\,,
\end{array} \right. \label{velocityrandom}
\eea
where we do not associate a turnover frequency to scales larger than the
correlation scale. Here $\om_k$ is the frequency of an eddy with
size $\ell<L$,
\bea
\omega_k &=& v_\ell/\ell=(v_L/L^{1/3})\ell^{-2/3} \nonumber\\
  &=& (v_L/L^{1/3})k^{2/3} \ll k, \quad k=1/\ell~.
\eea
The energy momentum tensor of the source in $\bk$
space involves a convolution (\cf section \ref{turbuanisostress}):
\be
T_{ij}(\bk,\eta)=\frac{4}{3}\rho(\eta) \int\frac{d^3q}{(2\pi)^3}\,
v_i(\bq)\,v_j(\bk-\bq)e^{i\omega_q \eta}e^{i\omega_{|\bk-\bq|} \eta}\,.
\ee
The convolution entails that the energy momentum
tensor as a random variable is oscillating also on large scales $k<L^{-1}$,
contrary to the random velocity field. At these scales the amplitude is
dominated by the contributions from $q=1/L$  (\cf
Eq.~(\ref{Pispectrum})).  We therefore expect the energy momentum
tensor to oscillate with the smallest frequency $\om_L$.  This can be
approximatively justified by the
following reasoning. Since we concentrate on the case of turbulence generated
by a first order phase transition, the source is active for a time $\tau_L$.
Therefore, the main contribution to the
convolution integral comes from frequencies such that $\omega_q \tau_L \sim 1$,
$\omega_{|\bk-\bq|}\tau_L \sim 1$. If $k<L^{-1}$ and $q<L^{-1}$ the
frequencies are both zero, but for $k<L^{-1}$ and $q>L^{-1}$ one can
approximate $\omega_{|\bk-\bq|}\sim \omega_q$ so that the frequency
is $2\omega_q$.  The same argument  for smaller scales
$k>L^{-1}$ gives the frequency $2\omega_{k}$.
We can therefore model the time dependence of the random anisotropic
stress tensor by
\bea
\Pi_{ij}(\bk,\eta)=\left\{ \begin{array}{ll}
\Pi_{ij}(\bk) e^{2i\omega_L \eta}& {\rm for}~k \leq L^{-1}\\
\Pi_{ij}(\bk) e^{2i\omega_k \eta} & {\rm for}~L^{-1}\leq k<\la^{-1}\,.
\end{array} \right. \label{anisotropicstessrandom}
\eea

With this source the gravitational wave equation becomes
the equation of a forced harmonic oscillator with damping,
\bea
\lefteqn{\ddot{h}_{ij}({\mathbf k},\eta)+\frac{2}{\eta}
\dot{h}_{ij}({\mathbf k},\eta)+k^2h_{ij}({\mathbf k},\eta)=} \label{gwequationforced}\\
& & = 8\pi G a^2(\eta)\Pi_{ij}({\mathbf k})e^{2i\omega \eta}
\Theta(\eta-\eta_{\rm in})\Theta(\eta_{\rm fin}-\eta)\nonumber\,.
\eea
Here the source frequency is generically denoted $2\omega$, and $\eta_{\rm
  in}$ and $\eta_{\rm fin}$ are the initial and final times of action of
the source, $\eta_{\rm fin}-\eta_{\rm in}=\tau_L$. The Heaviside
functions $\Theta$ are inserted to limit the source to the interval $\eta_{\rm
  in} < \eta < \eta_{\rm fin}$.
The source is active for a time much shorter than one
Hubble time: $\tau_L\ll \eta_*$. Therefore we can neglect the expansion of the
universe during the relatively short period during which the
source is active,
\bea
\lefteqn{\ddot{h}_{ij}({\mathbf k},\eta)+k^2h_{ij}({\mathbf k},\eta)=}
\label{gweqnactive}\\
& & = 8\pi G a^2_*\Pi_{ij}({\mathbf k})e^{2i\omega \eta}
\Theta(\eta-\eta_{\rm in})\Theta(\eta_{\rm fin}-\eta)\nonumber\,.
\eea
The solutions to the homogeneous equation without damping are simply
$\exp(\pm ik\eta)$, and with the Green's function method one finds the
inhomogeneous solution for $\eta \leq \eta_{\rm fin}$,
\bea
 h_{ij}(\bk,\eta\leq \eta_{\rm fin}) && =\frac{4\pi G a_*^2 \Pi_{ij}(\bk)}{k}
\times \nonumber\\
&&\hspace{-5mm}\left[e^{ik\eta}\frac{e^{i(2\omega-k)\eta}-e^{i(2\omega-k)\eta_{\rm
        in}}}{k-2\omega}+\right. \nonumber \\
&&\hspace{-3mm} \left.e^{-ik\eta}\frac{e^{i(2\omega+k)\eta}-e^{i(2\omega+k)\eta_{\rm in}}}{k+2\omega}
\right]\,. \label{solutionsourced}
\eea
Note that the singularity at $k=2\omega$ of the first term is removable.
For $\eta_{\rm in}<\eta<\eta_{\rm fin}$ the time dependence of
gravitational waves with wave number $\bk$ is given by the superposition
of the source frequency, $2\omega$, and of the source 
wavenumber $k$. However, once the source
ceases to be active, the gravitational wave propagates freely at the
speed of light. In order to find a solution for $\eta\gg \eta_{\rm
fin}$, we need to match Eq. (\ref{solutionsourced}) at $\eta_{\rm
fin}$ with a generic solution of the homogeneous propagation equation
\be
\ddot{h}_{ij}({\mathbf k},\eta)+\frac{2}{\eta}
\dot{h}_{ij}({\mathbf k},\eta)+k^2h_{ij}({\mathbf k},\eta)=0 \,.
\ee
Now the expansion of the universe can no longer be neglected, and the
homogeneous solutions in this case are $\exp(\pm ik\eta)/\eta$. The
matching procedure leads to
\bea
h_{ij}(\bk,\eta) &&=\frac{4\pi G a_*^2\eta_*
    \Pi_{ij}(\bk)}{k}\times \nonumber\\
&&\hspace{-6mm}
\left[\frac{e^{ik\eta}}{\eta}\frac{e^{i(2\omega-k)\eta_{\rm fin}}
-e^{i(2\omega-k)\eta_{\rm in}}}{k-2\omega}+ \right. \nonumber \\
&& \hspace{-3mm} \left.
\frac{e^{-ik\eta}}{\eta}\frac{e^{i(2\omega+k)\eta_{\rm fin}}
-e^{i(2\omega+k)\eta_{\rm in}}}{k+2\omega}\right]\label{solutionfinal}
\eea
(where we have substituted $\eta_{\rm fin}$ by $\eta_*$ in the
amplitude). For $\eta\gg \eta_*$ the gravitational wave propagates freely
with the correct dispersion relation: its frequency is just $\pm k$,
and the time dependence $\eta^{-1}$ accounts for the expansion of the
universe. The frequency $2\omega$ of the source only enters in the amplitude. 

We want to evaluate the gravitational wave energy density today, normalised to
the critical density $\rho_c$. In real space the  gravitational wave
energy density parameter is~\cite{michele}
\be
\Omega_G=\frac{\rho_G}{\rho_c} = \frac{\lag \dot{h}_{ij}
\dot{h}^{ij}\rag}{16\pi G \rho_ca^{2}}\,.
\ee
Here $\lag \cdots \rag$ denotes ensemble average and, at the same time, time
averaging over several periods of oscillation. The factor $1/a^2$ comes,
since the overdot denotes the derivative w.r.t. conformal time.

We define the gravitational wave power spectrum by
\be
\langle \dot{h}_{ij}(\bk,\eta) \dot{h}^{*}_{ij}(\bq,\eta)\rangle
   = \de(\bk-\bq) |\dot {h}|^2(k,\eta) \,.  \label{hk}
\ee
Fourier transforming the above relation for $\Omega_G$ one defines the
gravitational wave energy density per logarithmic unit of frequency:
\be
\Omega_G =\int\frac{dk}{k} \frac{d\Omega_G(k)}{d\log (k)}\,, \label{rhoG}
\ee
with
\be
\frac{d\Omega_G(k)}{d\log (k)} = \frac{k^3 |\dot h|^2}{4(2\pi)^6G\rho_c a^2}\,.
\label{OmegaG}
\ee
This is the quantity we are interested in. To proceed we need to find an expression for the
spectrum (\ref{hk}) at late times (today) and
on sub-horizon scales $k\eta\gg 1$. From solution (\ref{solutionfinal})
with $k\eta\gg 1$ one has, up to an irrelevant overall phase
\bea
\dot{h}_{ij}(\bk,\eta)&=&\frac{4\pi G a_*^2\eta_*
    \Pi_{ij}(\bk)}{\eta}\times \label{hdot}\\
&\times&\left[A e^{ik(\eta-\eta_{\rm in})}
- B e^{-ik(\eta-\eta_{\rm in})} \right]\,,\nonumber
\eea
where $A=\frac{e^{i(2\omega-k)\tau_L}-1}{k-2\omega}$ and
$B=\frac{e^{i(2\omega+k)\tau_L}-1}{k+2\omega}$.
Given the anisotropic stress power spectrum
(\ref{correlatorPi}), the gravitational wave power spectrum takes the form
\be
|\dot{h}|^2(k,\eta) =(4\pi G)^2 a^4_* \eta_*^2
\left(\frac{4\rho_*}{3}\right)^2 \Pi_v(k,\eta_*)
\frac{|A|^2+|B|^2}{\eta^2}\,.
\label{hdot2}
\ee

We now use Eqs. (\ref{OmegaG}) and (\ref{hdot2}) to evaluate the
gravitational wave energy density spectrum. The main spectral dependence comes from the
anisotropic stress spectrum (\ref{Pispectrum}), which we write in the form
\bea
k^3\Pi_v(k)\simeq \frac{27}{8}\pi^4
\left(\frac{\Om_T(\eta_*)}{\Om_\rad(\eta_*)}\right)^2
\left\{ \begin{array}{l}
\vspace*{0.2cm}
\frac{2}{13} x^3 \\
\text{for}~0<x<1  \vspace*{0.2cm} \\
 x^{-2/3} \\
  \text{for}~1<x<\frac{L}{\la}  \vspace*{0.2cm}\\
0 \qquad \text{otherwise}\,,
\end{array} \right.
\label{Pix}
\eea
where we have defined the variable $x=kL$. Moreover, the value of $\omega$
in the amplitudes $A$ and $B$ depends on $k$. 
From Eq.~(\ref{anisotropicstessrandom}), we have
$\om=\om_L$ for scales $k<L^{-1}$, and $\om=\om_k\propto k^{2/3}$ 
for scales $k>L^{-1}$. Using
the expression for the scale factor in the radiation dominated era
$a=H_0\sqrt{\Om_{\rm rad}}\eta$, we finally obtain the energy
density spectrum,
\begin{widetext}
\bea
\frac{d\Omega_G(k,\eta)}{d\log (k)}\simeq \frac{9}{64\pi}
\left(\mathcal{H}_*L\right)^2
\left(\frac{\Om_T(\eta_*)}{\Om_\rad(\eta_*)}\right)^2
\frac{\Om_{\rm rad}}{a^4(\eta)}\left\{ \begin{array}{ll}
\vspace*{0.2cm}
\frac{2}{13}\left[\frac{\left|e^{2i(1-\frac{x}{2v_L})}-1\right|^2}{(x-2v_L)^2}
+\frac{\left|e^{2i(1+\frac{x}{2v_L})}-1\right|^2}{(x+2v_L)^2}\right]
 \, x^3 & \text{for}~0<x<1\,,\\
\vspace*{0.2cm}
  \left[
 \frac{\left|e^{2i(x^{2/3}-\frac{x}{2v_L})}-1\right|^2}{(x-2v_Lx^{2/3})^2}
+\frac{\left|e^{2i(x^{2/3}-\frac{x}{2v_L})}-1\right|^2}{(x+2v_Lx^{2/3})^2}
\right] \, x^{-2/3}  & \text{for}~1<x<\frac{L}{\la} \\
0 & \text{otherwise}\,. \label{long}
\end{array} \right.
\eea
\end{widetext}

\begin{figure}
\begin{center}
\begin{minipage}{1\linewidth}
\centering
\epsfig{figure=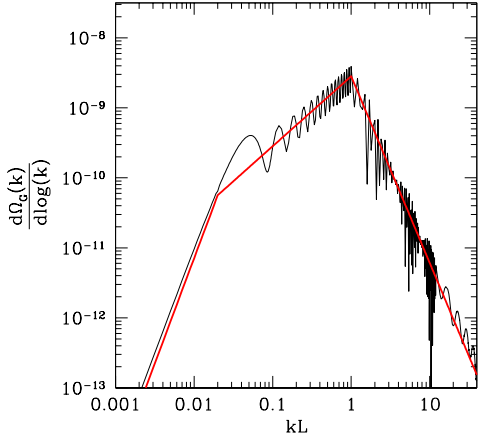,width=7cm}
\caption{The gravitational wave energy density today,
  normalised to $\left(\mathcal{H}_*L\right)^2 \Om_{\rm
  rad}$, as a function of $x=kL$ is shown. We have chosen
  $v_L^2=\frac{3}{2}\frac{\Om_T(\eta_*)}{\Om_\rad(\eta_*)}=10^{-4}$.
  The black, oscillating
  line shows the exact amplitude given in Eq. (\ref{long}). The red,
  straight line shows the approximated spectrum given in Eq. (\ref{Om0.01}),
  normalised to the same quantity $\left(\mathcal{H}_*L\right)^2 \Om_{\rm
  rad}$. The factor $2/13$ appearing in Eq. (\ref{long}) has been
  neglected in the plot since it generates an unphysical discontinuity.
\label{fig1} }
\end{minipage}
\end{center}
\end{figure}

\begin{figure}
\begin{center}
\begin{minipage}{1\linewidth}
\centering
\epsfig{figure=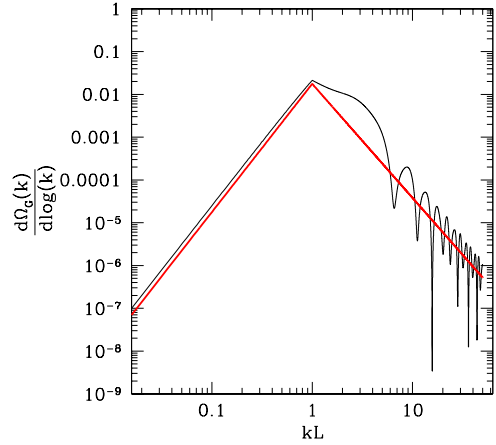,width=7cm}
\caption{Like Fig. \ref{fig1}, but with $v_L=1/2$, hence
  $\frac{\Om_T(\eta_*)}{\Om_\rad(\eta_*)} = 1/6$.
\label{fig2} }
\end{minipage}
\end{center}
\end{figure}

Once the source has decayed, the gravity wave energy density evolves
like radiation, as it should.
 The energy spectrum today is simply given by the above expression
evaluated at $\eta_0$ with $a(\eta_0)=1$. We obtain a
different spectral dependence
whether the largest eddy velocity $v_L$ is below or above
$1/2$. For $v_L\leq 1/2$, the characteristic frequency $2\omega_L$ is less than
the scale $L^{-1}$ at which the anisotropic stress power spectrum peaks
(\ref{Pix}). The gravitational wave energy spectrum still
peaks at $L^{-1}$, but it changes slope at
$k=2\om_L$. An approximate solution for the energy spectrum in the
case $v_L\leq 1/2$ is
\bea
&&\frac{d\Omega_G(k,\eta_0)}{d\log (k)}\simeq  \frac{9}{32\pi}
\left(\mathcal{H}_*L\right)^2
\left(\frac{\Om_T(\eta_*)}{\Om_\rad(\eta_*)}\right)^2
\Om_{\rm rad}
\times \nonumber \\
&&\times\left\{ \begin{array}{ll}
\vspace*{0.2cm}
 x^3/v_L^2 & \text{for}~0<x<2v_L\\
\vspace*{0.2cm}
4  \, x & \text{for}~2v_L<x<1\\
\vspace*{0.2cm}
4 \, x^{-8/3} & \text{for}~1<x<\frac{L}{\la}\\
0 & \text{otherwise}\,.
\end{array} \right. \label{Om0.01}
\eea
The value of the energy spectrum normalised to the parameters
$\left(\mathcal{H}_* L\right)^2 \Om_{\rm rad}$ is shown
in Fig.~\ref{fig1} for $v_L=0.01$ and in Fig.~\ref{fig2} for $v_L=1/2$.
We remind that $\frac{\Om_T(\eta_*)}{\Om_\rad(\eta_*)}=\frac{2}{3}v_L^2$ 
(\cf Eq.~(\ref{vL})).

\begin{figure}
\begin{center}
\begin{minipage}{1\linewidth}
\centering
\epsfig{figure=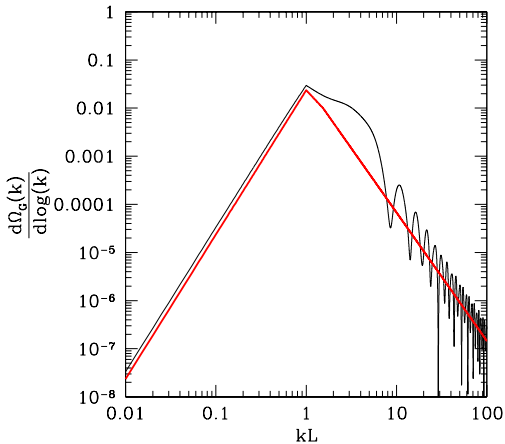,width=7cm}
\caption{The figure shows the gravitational wave energy density
  today, normalised to $\left(\mathcal{H}_*L\right)^2 \Om_{\rm
  rad}$, as a function of $x=kL$, for $v_L=1/\sqrt{3}$, hence
  $\frac{\Om_T(\eta_*)}{\Om_\rad(\eta_*)}=\frac{2}{9}$ . The red, 
  straight line is the approximation given in Eq. (\ref{Om3}).
\label{fig3}}
\end{minipage}
\end{center}
\end{figure}

If $v_L\geq 1/2$, then $2\om_k>2\om_L>L^{-1}$. The velocity of the largest
eddies is bounded by the speed of sound $v_L\leq 1/\sqrt{3}$
\cite{steinhardt,kosowsky}. 
Also in this case, the energy spectrum peaks at the
frequency $L^{-1}$, but because of the particular form of the amplitude it
changes slope at $k=(2 v_L)^3 / L$. An approximate solution for $v_L\geq 1/2$ is
\bea
&&\frac{d\Omega_G(k,\eta_0)}{d\log (k)}\simeq  \frac{9}{32\pi}
\left(\mathcal{H}_*L\right)^2
\left(\frac{\Om_T(\eta_*)}{\Om_\rad(\eta_*)}\right)^2
\Om_{\rm rad}
 \times \nonumber \\
&&\times\left\{ \begin{array}{ll}
\vspace*{0.2cm}
x^3/v_L^2 & \text{for}~0<x<1\\
\vspace*{0.2cm}
 x^{-2}/v_L^2 & \text{for}~1<x<(2v_L)^3\\
\vspace*{0.2cm}
4 \, x^{-8/3} & \text{for}~(2v_L)^3<x<\frac{L}{\la}\\
0 & \text{otherwise}\,.
\end{array} \right. \label{Om3}
\eea
This result is plotted in Fig. \ref{fig3} for the maximal value
$v_L=1/\sqrt{3}$, $\frac{\Om_T(\eta_*)}{\Om_\rad(\eta_*)}=\frac{2}{9}$.
Eqs.~(\ref{Om0.01}) and (\ref{Om3}) expressed in terms of the
parameters which describe the phase transition (see
Refs.~\cite{dolgov,nicolis}), can be found in Appendix~B.

The maximal gravitational wave signal is obtained for the highest
possible value of the eddy velocity, $v_L$.
Given that $L\simeq v_b(\al)\TT$, we obtain a peak frequency of
\be
k_L\equiv L^{-1}\simeq 8 \times 10^{-3}\frac{1}{v_b(\al)}
\frac{1}{\TT\mathcal{H}_*}\frac{T_*}{100\GeV} \,{\rm mHz}
\label{peakfre}
\ee
(we normalise it to the electroweak phase transition temperature).
Moreover, the phase transition has to be strongly
first order, so that the bubbles expand at the speed of light. 
If $v_b(\al)\simeq 1$ and $\TT\mathcal{H}_* \simeq 10^{-2}$ as argued
before, the spectrum peaks at a frequency of about mHz.
For the maximal eddy velocity of $v_L=1/\sqrt{3}$, we obtain the amplitude at the
peak frequency
\be
\left.\frac{d\Omega_G(k,\eta_0)}{d\log (k)}\right|_{k_L}\simeq
\frac{1}{24\pi}\times 10^{-4}\,\Omega_\rad \simeq
10^{-10}\,.
\ee
The total energy density in gravitational waves for $1/2\leq v_L\leq 1/\sqrt{3}$ is
given by the integral of Eq.~(\ref{Om3}),
\bea
\Om_G&=&\int_0^\infty \frac{dk}{k} \frac{d\Omega_G(k,\eta_0)}{d\log
  (k)}\nonumber\\ 
&\simeq& \frac{5}{32\pi}\left(\frac{L}{\eta_*}\right)^2
\frac{\Om_T(\eta_*)}{\Om_\rad(\eta_*)}\Om_\rad\,.
\label{totalturbu}
\eea
For the maximal value $v_L=1/\sqrt{3}$ we have therefore 
\bea
\Om_G &\simeq& \frac{5}{144\pi} \,\left(\frac{L}{\eta_*}\right)^2 \Om_{\rm rad}
\nonumber\\ 
&\simeq& 0.01 \, v^2_b(\al)
\left(\frac{\TT}{\eta_*}\right)^2 \Om_{\rm rad}. 
\eea
We find that the turbulent phase arising after a strongly first order
phase transition  produces a background of gravitational waves which
is in principle detectable by the space interferometer LISA. 
However, the gravity wave intensity obtained here depends
strongly on the peak amplitude of the turbulent spectrum, which is
rather uncertain. Therefore, also the amount of gravitational waves
generated is probably not very accurate.

Our result is in agreement with
Ref.~\cite{dolgov}, with a few differences. Since we
account for the natural dispersion relation of gravitational waves
$\om=k$, we obtain that the spectrum peaks at the frequency $1/L$, and
not at the  turnover frequency  $\om_L$ as found in
\cite{kamionkowski,kosowsky,dolgov}. The two differ by a factor $v_L$.
In our case the characteristic {\em size} of
the eddies determines the frequency of the induced gravitational waves, while
the eddy velocity enters only in the overall amplitude (see
Appendix~A). We therefore also have a different slope of the spectrum
for $k>L^{-1}$ than Refs.~\cite{kamionkowski,kosowsky,dolgov}. Moreover, 
we find that the gravitational wave spectrum is not reduced only to the `Kolmogorov'
part, but it continues to wavelengths larger than $1/L$ as shown in
Figs~\ref{fig1} to \ref{fig3}. For a strongly first order phase transition,
the amplitude of the signal at mHz frequencies derived here is of the same
order of magnitude as found in \cite{dolgov}, whithin the range of our
analytical approximations.

\section{Gravitational waves from magnetic fields}
\label{magnetic}

\subsection{The magnetic field power spectrum}

We now analyse the production of gravitational waves by a
primordial magnetic field. 
In order to make a direct comparison with the turbulence case, we consider
here only causally created magnetic fields.
Typically, like turbulence, magnetic fields can be generated during a phase
transition (note however that in the case of the electroweak phase
transition, magnetic seed fields can form even if the phase transition
is second order~\cite{magcausalgeneration}).

The magnetic field power spectrum on large scales is determined along
the same lines as the turbulence power spectrum
in Section~\ref{powerspectrum}. The field is divergence-free
and it must be uncorrelated above a certain scale,
therefore the slope has to be $k^n$ with $n\geq 2$ an even
integer. This slope continues up to the correlation scale. If the
magnetic field is simply constant on scales smaller 
than the correlation scale the power spectrum goes to zero steeply on
these scales (in \cite{hind} for example it is modeled as
$\exp(-(k/k_c)^4)$). This is normally assumed to be the initial
spectrum. However as the system evolves, the interactions 
of the magnetic field with the cosmic fluid modify the spectrum at small
scales, turning it into a power law (for example, into the
Iroshnikov Kraichnan spectrum, $E_B(k)=C(\varepsilon
v_A)^{1/2} k^{-3/2}$ with $v_A$ the Alfven speed, see
\cite{kosowsky,dolgov,hind,banj}).
To stay general, we make the ansatz
\be
\lag B_{i}(\bk)B_j^*(\bq)\rag = (2\pi)^3
\de(\bk-\bq)(\de_{ij}-\hat{k}_i\hat{k}_j) P_B(k)\,,
\ee
\bea
P_B(k)=\left\{ \begin{array}{ll}
C_B \,k^2 &\text{for}~0<k<L^{-1}\\
C_B \,L^{\al-2}k^{\al} &\text{for}~L^{-1}<k<\la^{-1}\\
0 &\text{otherwise}\,,
\end{array} \right. \label{magspectrum}
\eea
with $\al<-3$, so that the integrated energy is
dominated by the contribution from the correlation scale $L$.
$L$ and $\la$ are not necessarily the same scales as for turbulence but are
the magnetic correlation and dissipation scales. $C_B$ is an as yet
undetermined constant.

Once the magnetic field is generated, it is not quickly
dissipated like turbulence. Because of the
high conductivity of the cosmic plasma there is negligible Ohmic dissipation
and the magnetic flux is conserved. During the early stages of
evolution of the universe both, the kinetic and magnetic Reynolds
numbers are very high, and 
magnetohydrodynamical turbulence is generated. The magnetic field decay
has been analysed in this case both analytically~\cite{olesen,shiro,son,campa} and
numerically \cite{hind,banj}, and different scaling laws are obtained
depending on the treatment of the problem. In the following we keep the
scaling exponent as a free parameter, since it will not be very relevant for
the final result. However, all papers agree on the fact that the power
spectrum is persistent on large scales, {\em i.e.} that $C_B$ is
constant in time.

After electron-positron annihilation
the neutrinos are already decoupled and the photon
mean free path increases suddenly . Then the system
is no longer turbulent but dissipative due to radiation viscosity,
and the magnetic field evolution changes~\cite{jeda,subra,banj}.

The main point for our analysis is that
in the process of generation of gravitational
waves, one needs in principle to account for the evolution of the magnetic
spectrum. This can influence the final result, since the magnetic
field is acting as a source of gravitational waves for a
cosmologically long time. The situation is different from the case of
turbulence. This evolution has not been
taken into account in \cite{nucleosintesi}, where a fixed correlation
length given by the size of the horizon at the time of generation has
been assumed.

As for the slopes of the magnetic spectrum (\ref{magspectrum}), we
make a general 
Ansatz also for its time evolution. The evolution is mainly determined
by three processes: the expansion of the universe, the growth of the
correlation length and the decay of the energy density. To eliminate
the time dependence by redshifting we work with the comoving magnetic energy
density
\be
\lag B^2\rag =\lag B^2 (\eta)\rag a^4(\eta)\,.
\label{Bcomov}
\ee
 However, as we shall see below, $\lag B^2\rag$ is still time
 dependent due to dissipation of energy.

As in the case of the turbulence, we normalise the spectrum
(\ref{magspectrum}) in terms of the total magnetic energy density divided by
the radiation energy density:
\bea
\frac{\Om_B(\eta)}{\Om_\rad(\eta)}
  &=& \frac{1}{\rho_{\rad}(\eta_0)}\frac{\lag B^2\rag}{8\pi}  
\nonumber \\  &=&
\frac{1}{\rho_{\rad}(\eta_0)}\frac{1}{(2\pi)^3}\int_0^\infty
\frac{dk}{k}k^3P_B(k)\nonumber \\ 
&\equiv& \frac{1}{\Om_\rad}\int_0^\infty\frac{dk}{k} \frac{d\Om_B}{d\log k}
\eea
which gives
\bea
C_B &\simeq&\frac{(2\pi)^3}{8\pi}\frac{5(\al+3)}{\al-2}\lag B^2\rag L^5 \label{CB}\\
&=& (2\pi)^3\frac{5(\al+3)}{\al-2}\Om_B\rho_c L^5 \,. \nonumber
\eea
Here we have defined the magnetic field energy density parameter $\Om_B$
which corresponds to the magnetic field energy density scaled to today
via Eq.~(\ref{Bcomov}). $\Om_B$
is itself a function of time due to the dissipation of energy.
%

Let us now study the time dependence of  $L$ and $\lag B^2\rag$. We
model the comoving correlation length by 
\be
L(\eta)=L_{*}\left(\frac{\eta}{\eta_*}\right)^\ga\,,
\ee
with $0<\ga<1$ so that $L(\eta)$ never overcomes the horizon at any instant
$\eta$. $L_*$ denotes the correlation length at the 
epoch $\eta_*$ of creation of the magnetic field.
The time dependence of the dissipation scale $\la$ does not play a significant
role in our analysis, so we do not consider it. For our purposes, it
is enough to have in 
mind that $\la$ is also a function growing with time. 
In order to find the time evolution of the magnetic energy density,
we make an analogy: we assume that from the
MHD equation it is possible to define a conserved quantity analogous to the
Loitsyansky's integral. We impose simply
\be\label{conserv}
\lag B^2\rag L^5 ={\rm constant~in~time,}
\ee
with this the large scale part of the spectrum is persistent (see
Eq.~(\ref{CB}) above). This is equivalent to the constancy of $C_B$,
which has been observed in simulations~\cite{hind,banj}. 
With this the comoving magnetic energy scales as
\be\label{B^2a^4}
\lag B^2\rag(\eta)=\lag B^2\rag(\eta_{*})
\left(\frac{\eta_{*}}{\eta}\right)^{5\ga}\,.
\ee
If we were to adopt the same decay laws as in turbulence, one would find the
Kolmogorov decay laws $L(\eta)\propto \eta^{2/7}$, $\lag B^2\rag(\eta)\propto
\eta^{-10/7}$. Equivalent scaling laws are obtained from the
arguments of self similarity used in \cite{olesen,hind}, and of
selective decay used in
\cite{son,banj}. The main difference is that most of these works consider
a white noise spectrum for the magnetic field at large scales, resulting in a
slower decay of the energy and a more substantial growth of the
correlation length. They find Saffman's law for fluid dynamics. However,
as demonstrated in Section~\ref{powerspectrum}, causally generated,
divergence-free vector fields in the early
universe necessarily lead to the formation of a Batchelor
spectrum. This is valid for both, turbulence and magnetic fields.

\subsection{Power spectrum of the anisotropic stress}

From the (comoving) magnetic field stress tensor
\be
T^{(B)}_{ij} =\frac{1}{4\pi}\left[\frac{1}{2}B^2g_{ij} -B_i B_j\right]
\ee
one defines again the anisotropic stress as
$\Pi_{ij}(\bk)=(P_{il}P_{jm}-1/2P_{ij}P_{lm})T_{lm}(\bk)$.
One then calculates the comoving anisotropic stress power spectrum in the
usual way, see \eg \cite{RPT}, using Wick's theorem to reduce 4-point
to 2-point correlators under the assumption of Gaussianity (which
again is most probably not strictly valid but can be expected to give
the right order of magnitude). Setting
\be
 \lag\Pi_{ij}(\bk,\eta)\Pi^*_{ij}(\bq,\eta)\rag \equiv
\Pi_B(k,\eta)\de(\bk-\bq)
\ee
one obtains, exactly like in Eq.~(\ref{Pispectrumint}),
\be
\Pi_B(k,\eta)=\int d^3p P_B(p,\eta)P_B(|\bk-\bp|,\eta)
(1+\gamma^2)(1+\beta^2)\, .
\ee
A good approximation to this convolution integral is
\bea
\Pi_B(k,\eta)\simeq C_B^2 \left\{  \begin{array}{l}
A_1^2\,L^{-7}(\eta) \\  \text{for}~0<k<L^{-1} \vspace*{0.2cm}\\
A_2^2\,L^{\al-7}(\eta)\,k^{\al} \\
 \text{for}~L^{-1}<k<\la^{-1} \vspace*{0.2cm}\\
0 \quad \text{otherwise}\,,
\end{array} \right. \label{maganisostress}
\eea
with $A_1=\sqrt{\frac{2\al-4}{7(2\al+3)}}$ and $A_2=\sqrt{\frac{\al-2}{5(2\al+3)}}$.

The time evolutions of the magnetic energy density and of the
correlation scale are deterministic. The same holds for the
evolution of the anisotropic stress power spectrum. We therefore may treat
magnetic fields as a coherent source, \ie a source with deterministic
time dependence,
\bea
\lefteqn{\lag\Pi_{ij}(\bk,\eta_1)\Pi^*_{ij}(\bq,\eta_2)\rag =} \nonumber\\
& f(\eta_1)f(\eta_2)
\lag\Pi_{ij}(\bk,\eta_\mr{in})\Pi^*_{ij}(\bq,\eta_\mr{in})\rag\de(\bk-\bq)\,,
\eea
($f(\eta)$ denotes here an arbitrary function). 
Using $\Pi_B(\bk,\eta)=f^2(\eta)\Pi_B(\bk,\eta_{\rm in})$, we find
\be
\lag\Pi_{ij}(\bk,\eta_1)\Pi^*_{ij}(\bq,\eta_2)\rag =
\sqrt{\Pi_B(k,\eta_1)} \sqrt{\Pi_B(k,\eta_2)}\de(\bk-\bq)\,.
\label{coherence}
\ee

\subsection{Generation of gravitational waves}

In the case of a coherent source, one can write the evolution
equation (\ref{gwevolution}) directly for the gravitational wave power
spectrum. To see this, we first write it for the stochastic comoving
anisotropic stress of the magnetic field
\be
{\ddot{h}}_{ij}+\frac{2}{\eta}\dot{h}_{ij}+
k^2 h_{ij}=8\pi G \frac{\Pi_{ij}}{a^2}\,. \label{gweq}
\ee
The solution to this equation with initial condition $h_{ij}(\eta_*)=0$
 is
\be
h_{ij}(\bk,\eta)=8\pi G\int_{\eta_*}^{\eta}d\zeta\,
\frac{\Pi_{ij}(\bk,\zeta)}{a^2(\zeta)}\mathcal{G}(\eta,\zeta,\bk)\,,
\ee
where $\mathcal{G}(\eta,\zeta,\bk)$ is the corresponding Green
function. The gravitational wave energy power spectrum becomes
\bea
&&\lag \dot h_{ij}(\bk,\eta)\dot h_{ij}^*(\bq,\eta)\rag=(8\pi G)^2
\frac{d}{d\eta}\int_{\eta_*}^{\eta}d\zeta
\frac{\mathcal{G}(\eta,\zeta,\bk)}{a^2(\zeta)} \nonumber\\
&&\times\frac{d}{d\eta}\int_{\eta_*}^{\eta}d\xi
\frac{\mathcal{G}^*(\eta,\xi,\bq)}{a^2(\xi)}
\lag \Pi_{ij}(\bk,\zeta) \Pi_{ij}^*(\bq,\xi)\rag\,.
\eea
For a coherent source, using (\ref{coherence}), we find
\bea
\lefteqn{\lag \dot h_{ij}(\bk,\eta)\dot h_{ij}^*(\bq,\eta)\rag
=|\dot h|^2(k,\eta)\, \de(\bk-\bq)=} \\
&=&\de(\bk-\bq)\left|8\pi G \frac{d}{d\eta}\int_{\eta_*}^{\eta}
d\zeta \frac{\sqrt{\Pi_B(k,\zeta)}}{a^2(\zeta)}
\mathcal{G}(\eta,\zeta,\bk)\right|^2 \nonumber
\eea
where we can identify
\be
h(k,\eta)=8\pi G\int_{\eta_*}^{\eta}d\zeta
\frac{\sqrt{\Pi_B(k,\zeta)}}{a^{2}(\zeta)}
\mathcal{G}(\eta,\zeta,\bk)\,.
\ee
Since
$\dot{h}_{ij}(\bk,\eta)$ and $h(k,\eta)$ share the same Green's
function, $h(k,\eta)$ satisfies
equation (\ref{gweq}), with source term $8\pi G
\sqrt{\Pi_B(k,\eta)}/a^2$. In the radiation era this is
\be
\ddot h+\frac{2}{\eta}\dot h +
k^2h=\frac{8\pi G}{H_0^2\Omega_{\rm rad}\eta^2}\sqrt{\Pi_B(k,\eta)} ~.
\label{Hprime}
\ee
With the power spectrum (\ref{maganisostress}), the solution is
\bea
h(k,\eta)&=&\frac{4\pi G}{H_0^2\Omega_{\rm rad}}\, \frac{i}{k}\, C_B
\nonumber \\
& & \int_{\eta_*}^\eta  d\zeta
\left[ \frac{e^{-ik\eta}}{\eta}\frac{e^{ik\zeta}}{\zeta}-
\frac{e^{ik\eta}}{\eta}\frac{e^{-ik\zeta}}{\zeta}\right]\times \nonumber\\
& &\left\{  \begin{array}{l}
A_1\,L^{-7/2}(\zeta) \\ \text{for}~0<k<L^{-1}(\zeta) \vspace*{0.2cm} \\
A_2\,L^{(\al-7)/2}(\zeta)\,k^{\al/2}\\
  \text{for}~L^{-1}(\zeta)<k<\la^{-1}(\zeta) \vspace*{0.2cm} \\
0 \qquad \text{otherwise}\,.
\end{array} \right. \label{hmagnetic}
\eea
For a given mode $k$, the time integration has to be divided in different
ranges: if the mode satisfies initially $k<L_*^{-1}$, then there exists a time
$\eta_L(k)$ at which the correlation length overcomes $k^{-1}$, and we
have $k\geq 
L^{-1}(\eta)$ for $\eta\geq \eta_L(k)$. From this time on, one has to consider
the small scale part of the source spectrum. Moreover, the modes 
that are super-horizon at beginning $k\leq \eta_*^{-1}$
have to be considered separately, since in this case the integrand does not oscillate 
until horizon crossing. 
However, since the magnetic correlation length $L(\eta)$ is increasing with
time, the above integrals are in any case dominated by their values at the lower
bound $\eta_*$.  We split the solution in three $k$
regions, and we finally obtain
\bea
&h(k,\eta)&\simeq\frac{4\pi G}{H_0^2\Omega_{\rm rad}}\, \frac{i}{k}\, C_B
\nonumber \\
&&\left\{  \begin{array}{l}
A_1\,L_*^{-7/2} \frac{2}{7\ga}
\left[ \frac{e^{-ik\eta}}{\eta}-\frac{e^{ik\eta}}{\eta}\right]
\\ \text{for}~0<k<\eta_*^{-1} \vspace*{0.2cm} \\
A_1\,L_*^{-7/2}\frac{i}{k\eta_*} 
\left[ \frac{e^{-ik(\eta-\eta_*)}}{\eta}+\frac{e^{ik(\eta-\eta_*)}}{\eta}\right]
\\ \text{for}~\eta_*^{-1}<k<L_*^{-1} \vspace*{0.2cm} \\
A_2\,L_*^{-7/2}(k L_*)^{\al/2}\frac{i}{k\eta_*} 
\left[ \frac{e^{-ik(\eta-\eta_*)}}{\eta}+\frac{e^{ik(\eta-\eta_*)}}{\eta}\right]
\\ \text{for}~L_*^{-1}<k<\la_*^{-1} \vspace*{0.2cm} \\
0 \qquad \text{otherwise}\,.
\end{array} \right. 
\eea
This solution has to be matched with the solution of the homogeneous equation  
at the time the source ceases to be active: that is, for every mode $k$, the
time $\eta_\la(k)$ at which the dissipation scale has grown bigger than
$k^{-1}$. However, the above solution does not depend on $\eta_\la(k)$, being
dominated by the lower bound $\eta_*$, and has the same time
dependence as a free propagating wave. Therefore, the above solution
is simply valid also for $\eta>\eta_\la(k)$. 

\begin{figure}
\begin{center}
\begin{minipage}{1\linewidth}
\centering
\epsfig{figure=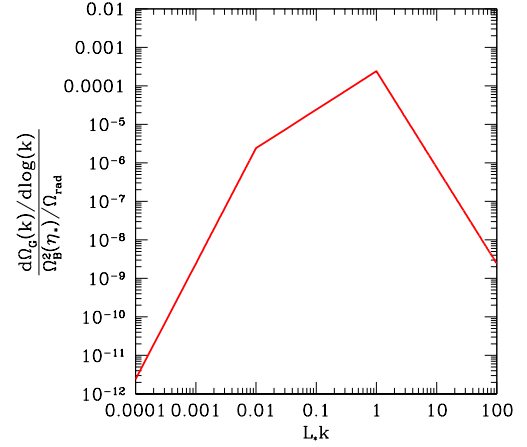,width=7cm} \\
 \epsfig{figure=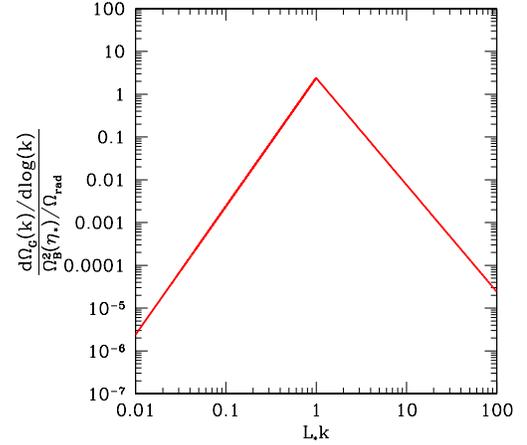,width=7cm}
\caption{The gravitational wave energy density today,
  normalised to $\Om_B^2/\Om_{\rm
  rad}$, as a function of $x=kL_*$ is shown for $L_*\HH_*=0.01$ in the
  top panel and for $L_*\HH_*=1$ in the bottom panel. On
  scales $k<1/\eta_*$ the spectrum behaves as $k^3$, then it flattens
  into linear growth up to the correlation scale $L_*^{-1}$. On
  smaller scales it decays like $k^{\al+1}$. In this plot $\al = -7/2$
  and $\ga=2/7$ are chosen.  
\label{fig4} }
\end{minipage}
\end{center}
\end{figure}

We want to evaluate the
gravitational wave energy density per logarithmic unit of
frequency given in Eq.~(\ref{OmegaG}). For this we have to average
$|\dot h|^2$ over several periods. Substituting the expression of the scale
factor $a(\eta)=H_0\sqrt{\Om_\rad} \eta$ we find
\bea
&|\dot h|^2(k,\eta)&=\frac{(4\pi G)^2}{H_0^2\Omega_{\rm rad}}\, C_B^2\,
\frac{2}{a^2(\eta)}
\nonumber \\
& &\left\{  \begin{array}{l}
A_1^2\,L_*^{-7} \big(\frac{2}{7\ga}\big)^2
\,\,\, \text{for}~0<k<\eta_*^{-1} \vspace*{0.2cm}\\
A_1^2\,L_*^{-7}\big(\frac{1}{k\eta_*}\big)^2
\,\,\, \text{for}~\eta_*^{-1}<k<L_*^{-1} \vspace*{0.2cm}\\
A_2^2\,L_*^{-7}(kL_*)^{\al} \big(\frac{1}{k\eta_*}\big)^2
\,\,\, \text{for}~L_*^{-1}<k<\la_*^{-1}  \vspace*{0.2cm}\\
0 \qquad \text{otherwise}\,.
\end{array} \right. 
\eea
Here we have used that
$\dot h \simeq kh$ on sub-horizon scales. The gravitational wave
energy density is now given by (\ref{OmegaG}). We substitute the constant 
$C_B$ using expression (\ref{CB}), and since $\Om_B L^5$ is time independent,
we can insert its value at the initial time,
$\Om_B(\eta_*)L_*^5$. These manipulations lead to (see fig. \ref{fig4})
\bea
&\frac{d\Omega_G(k,\eta_0)}{d\log (k)}&\simeq 75 \pi^2 
\left(\frac{\al+3}{\al-2}\right)^2 \frac{\Om_B^2(\eta_*)}{\Om_\rad}
\nonumber \\
& &\left\{  \begin{array}{l}
A_1^2\, \big(\frac{2}{7\ga}\big)^2 (L_* k)^3
\,\,\, \text{for}~0<k<\eta_*^{-1} \vspace*{0.2cm} \\
A_1^2\,\big(\frac{L_*}{\eta_*}\big)^2 (L_* k)
\,\,\, \text{for}~\eta_*^{-1}<k<L_*^{-1} \vspace*{0.2cm} \\
A_2^2\,\big(\frac{L_*}{\eta_*}\big)^2 (L_* k)^{\al+1}
\,\,\, \text{for}~L_*^{-1}<k<\la_*^{-1} \vspace*{0.2cm} \\
0 \qquad \text{otherwise}\,.
\end{array} \right. \label{GWspectrumB}
\eea

The total energy density can be approximated by 
\bea
&& \Om_G  =  \int_0^{\infty} \frac{dk}{k}
\frac{d\Omega_G(k,\eta_0)}{d\log (k)}\\
&&\simeq 75 \pi^2 
\left(\frac{\al+3}{\al-2}\right)^2 \left( A_1^2 -\frac{A_2^2}{\al+1}\right)
\left(\frac{L_*}{\eta_*}\right)^2
\frac{\Om_B^2(\eta_*)}{\Om_\rad}\nonumber\,,
\eea
where we have neglected a term $(L_*/\eta_*)^3$, given that $L_*\leq \eta_*$. If
the magnetic field at small scales develops an Iroshnikov Kraichnan spectrum
corresponding to $\al=-7/2$, the above quantity becomes
\be
\Om_G \simeq \frac{24}{77} \pi^2 
\left(\frac{L_*}{\eta_*}\right)^2
\frac{\Om_B^2(\eta_*)}{\Om_\rad}\,.
\label{Omegatotmag}
\ee

\subsection{Limits on the magnetic field intensity}

In paper \cite{nucleosintesi}, we found strong constraints on the
amplitude of a 
primordial magnetic field by applying the nucleosynthesis bound on the
gravitational waves generated by the magnetic field. The limits apply to 
magnetic fields created both during inflation and during a phase
transition, and are formulated in terms of the magnetic field
intensity $B_\La$ present today on the scale $\La= 0.1 \Mpc$. This is
the relevant scale for the
fields observed in clusters of galaxies. Since the magnetic 
field generation takes place long before $\eta=0.1\Mpc\simeq
10^{13}\sec$,  the  mode $\La$ is  super-horizon at
the moment of creation. 

If the magnetic
field is generated at inflation, it is characterised 
by a simple power law spectrum $k^n$ with $n>-3$, up to an upper cutoff which
corresponds to the time at which inflation ends, $k<\eta_*^{-1}$, 
where $\eta_*\simeq 8\times 10^{-9}\sec$ is a typical value. The
generation of magnetic fields during a phase 
transition instead is a causal process, leading mainly to fields on scales
smaller than the size of the horizon at the moment the field is
created. In the case of the electroweak phase transition at $100\GeV$,
this corresponds to $\eta_*\simeq  10^{5}\sec$. 

In Ref.~\cite{nucleosintesi} we have neglected sub-horizon fields,
which cannot propagate to larger scales unless they are helical and
subject to an inverse cascade. Motivated by inflation,  we
have imposed a cutoff scale $\eta_*^{-1}$ in the magnetic power
spectrum corresponding to the horizon size 
at the phase transition.  We have argued, that this leads to
conservative bounds for the magnetic fields. Here we show that
taking into account also the sub-horizon contribution
strengthens the limits somewhat. 

As we have seen, gravitational wave production takes place also inside
the horizon. Moreover, under realistic circumstances, most of the
magnetic energy is stored on these scales. Therefore, we now
recalculate the bounds obtained in~\cite{nucleosintesi} accounting also for
sub-horizon modes. The results we find below reduce to those of
Ref.~\cite{nucleosintesi} in the limiting case $L_* \simeq\eta_*$. As for the
previous sections, our results reside on the assumption that the large scale
part of the magnetic spectrum is persistent. 

The nucleosynthesis limit on any additional radiation-like form of energy
comes from the fact that this latter may not significantly change the
expansion law of the universe during nucleosynthesis. The maximum allowed
additional energy density redshifted to today is about $\Om_N\lsim
10^{-5}$ \cite{michele}. 
We impose this bound on the gravitational wave energy density generated by the
magnetic field. To be specific, we now fix the value $\al=-7/2$, but
this choice does not affect the final result in a relevant way. From
Eq.~(\ref{Omegatotmag}), we have therefore
\be
\Om^2_B(\eta_*)\lsim \frac{77}{24\pi^2}\times 10^{-5}\, 
\Om_\rad \left(\frac{\eta_*}{L_*}\right)^2\,.
\label{omegalim}
\ee
The upper bound on the total magnetic energy density is increased
for $ L_*\leq \eta_*$. We want to formulate the bound in
terms of the magnetic field amplitude on the scale $\La=0.1\Mpc$. At this aim,
we perform a volume average of the field in a region of size $\La^3$. 
We convolve the field with a Gaussian window function
\be
\bB_\La(\bx)=\int \frac{d^3 y}{V_\La}\,
\bB(\bx+\by)\exp\left(-\frac{y^2}{2\La^2}\right)\,,
\ee
and we define the magnetic field energy density smoothed on a scale $\La$, 
$B_\La^2=\lag \bB_\La(\bx)\cdot \bB_\La(\bx)\rag$. We have therefore
\be
B_\La^2=\frac{1}{\pi^2}\int_0^\infty dk k^2P_B(k)\exp\left(-\frac{\La^2 k^2}{2}\right)\,.
\ee
Since $\La\gg L_*$, the above integral is dominated by the large scale part of
the spectrum $kL_*\leq 1$, and inserting Eq.~(\ref{CB}) we obtain finally 
\be
B_\La^2\simeq \frac{120}{11}\sqrt{2\pi}\,\,\rho_c\,\Om_B(\eta_*) 
\left(\frac{L_*}{\La}\right)^5\,,
\ee
were again we have taken $\Om_B L^5$ at the initial time. 
We can now use Eq.~(\ref{omegalim}) to derive an upper bound on the
magnetic field intensity $B_\La$:
\be
\frac{B_\La}{10^{-6}\Gauss}\lsim \left(\frac{L_*}{\eta_*}\right)^2 
\left(\frac{\eta_*}{\La}\right)^{5/2}\,.
\ee
Comparing with Eq.~(33) of Ref.~\cite{nucleosintesi}, we see that the bound is
strengthened by a factor $\big(\frac{L_*}{\eta_*}\big)^2$. The reason for
that is twofold. First, the limit on $\Om_B$ is increased by a factor
$\eta_*/L_*$  when
reducing $L_*$ for fixed $\eta_*$. On the other hand, the magnetic
field is peaked at the smaller scale $L_*$: this appears in the value
of $B_\La^2$ through the factor $(L_*/\La)^5= (L_*/\eta_*)^5(\eta_*/\La)^5$. 

In the case of the electroweak phase transition at $100\GeV$, 
for which $\eta_*\simeq 10^{5}\sec$, we find
\be
B_\La\lsim 10^{-26}\left(\frac{L_*}{\eta_*}\right)^2\Gauss\,.
\ee
If the phase transition is first order and proceeds via bubble
nucleation, we expect the correlation length to be given by the size
of the largest bubbles which has been estimated as $L_*\sim
0.01\eta_*$, leading to the gravity wave spectrum shown in the top
panel of Fig.~\ref{fig4}. However, if the phase transition is second order, the
magnetic field inside one horizon volume is created smoothly and has
enough time to align itself. In this case we therefore expect a
correlation length of the order of $L_*\simeq \eta_*$, leading to the
gravity wave spectrum shown in the bottom 
panel of Fig.~\ref{fig4}.

\section{Comparison of the two cases}
\label{comparison}

We have calculated gravity wave production from two sources which are
both represented by divergence free vector fields. One of them (the
magnetic field) is persistent over many Hubble times while turbulence
decays within a fraction of one Hubble time. We would naturally expect the
former to produce more gravity waves than the latter. However, this is not
properly the case. Here we compare
the efficiency of the two sources, turbulent and magnetic, in detail.
Let us first compare the total
gravitational wave energy produced, as a function of the total energy
available in the sources. For this, we consider
Eqs.~(\ref{totalturbu}) and (\ref{Omegatotmag}), where the 
first one applies for the largest eddy velocity $1/2\leq v_L \leq 1/\sqrt{3}$
and the second one for the value of the magnetic field spectrum
$\al=-7/2$. Denoting the quantities concerning turbulence by a
superscript $T$ and those concerning magnetic fields by $B$,
we have
\be
\frac{\Om_G^T}{\Om_G^B}\simeq 0.02 \left( \frac{L_*^T}{L_*^B}\right)^2
\frac{\left(\frac{\Om_T(\eta_*)}{\Om_{\rad}(\eta_*)}\right)}
{\left(\frac{\Om_B(\eta_*)}{\Om_{\rad}}\right)^2}\,.
\label{compatot}
\ee
The gravitational wave energy density generated by magnetic fields depends
quadratically on the total energy of the magnetic source, while for
turbulence the relation is linear. This is due to the presence of the factor
$v_L^{-2}=\frac{2}{3}\frac{\Om_\rad(\eta_*)}{\Om_T(\eta_*)}$ in the 
amplitude of the gravitational wave turbulent spectrum~(\ref{Om3}). 
If the turbulent and magnetic correlation scales as well as the total
energy available in the sources are comparable, the efficiency in generating 
gravitational waves is similar. For
$v_L=1/\sqrt{3}$, we find from Eq.~(\ref{compatot}) 
$\Om_G^T/\Om_G^B\simeq 0.1$, roughly the same order
of magnitude within our accuracy. We find no significant domination of the
magnetic field induced total gravity wave energy. 

We now compare the gravitational wave energy density spectra, in
order to analyse the efficiency of the sources as a function of wavenumber $k$.  
We want to relate Eqs.~(\ref{Om3}) and (\ref{GWspectrumB}). For this
we set the correlation lenghts to the same value $L_*^T=L_*^B=L_*$. 
Therefore we take $L_*\mathcal{H}_*\simeq 0.01$, as in the case of turbulence
generated after a phase transition (\cf section \ref{turbulence}).
Moreover, we further impose that the initial available energy densities
are comparable, $\Om_T(\eta_*)/\Om_{\rad}(\eta_*)\simeq 
\Om_B(\eta_*)/\Om_{\rad}$.
Applying the reasonable values $v_L=1/\sqrt{3}$, $\al=-7/2$,
$\ga=2/7$, the ratio of Eqs.~(\ref{Om3}) and (\ref{GWspectrumB})
becomes ($x=kL_*$) 
\bea
\frac{\left(\frac{d\Omega_G^T(k,\eta_0)}{d\log (k)}\right)}
{\left(\frac{d\Omega_G^B(k,\eta_0)}{d\log (k)}\right)} \simeq 
0.1
\left\{  \begin{array}{l}
(L_* \mathcal{H}_*)^2
\,\,\, \text{for}~0<x<L_* \mathcal{H}_* \vspace*{0.2cm} \\
x^2
\,\,\, \text{for}~L_* \mathcal{H}_*<x<1 \vspace*{0.2cm} \\
\frac{10}{7}\,x^{1/2}
\,\,\, \text{for}~1<x<(2v_L)^3 \vspace*{0.2cm} \\
\frac{40}{21}\,x^{-1/6}
\,\,\, \text{for}~(2v_L)^3<x<L_*/\la_* \vspace*{0.2cm} \\
0 \,\,\, \text{otherwise}\,.
\end{array} \right. 
\eea
From the above expression we can conclude that, on super-horizon wavelenghts
$x<L_* \mathcal{H}_*$, $k<\eta_*^{-1}$, the energy density of 
gravitational waves from turbulence is smaller than the one
from magnetic fields by a factor $0.1(L_*\mathcal{H}_*)^2 \simeq
10^{-5}$. For sub-horizon modes, the ratio is increasing and reaches the
maximum value at $k_{\max}=(2v_L)^3/L_*$, where the two amplitudes 
are comparable  
$\frac{d\Omega_G^T(k,\eta_0)}{d\log(k)}/\frac{d\Omega_G^B(k,\eta_0)}{d\log (k)}
\simeq 0.1$.  Since the spectrum peaks at $k=1/L_*\simeq k_{\max}$, also
the total energy densities are comparable.

We hence find that on super-horizon scales magnetic fields are
more efficient than turbulence in generating gravitational waves. 
This is due to the difference of the time
intervals over which the  sources act, and to the characteristic way
in which a radiation-like source converts energy into gravitational
waves. In the integrals of  
Eq.~(\ref{hmagnetic}), the Green's function oscillates with frequency
$k$ and decays as a power law. Therefore,  super-horizon  modes,
$k\eta\leq 1$  are more efficiently converted into gravitational waves
than sub-horizon modes. Since the magnetic field is active for many
Hubble times (approximately from its generation up the epoch of
matter-radiation equality), the  
magnetic super-horizon modes generate gravity waves until they cross
the horizon (subsequent generation can be neglected). Turbulence, on
the other hand, is active as a source of gravitational waves only for a
time interval much shorter than one Hubble time $\tau_L\ll\HH_*^{-1}$. Most
turbulent modes that start super-horizon do not have the time to convert their
energy in gravitational waves. This leads to the suppression factor
$(L_*/\eta_*)^2 =(L_*\HH_*)^2 $ for the gravity wave spectrum from turbulence.

The expression for the turbulence to magnetic field ratio in full
generality (with $L_* =L^T=L_*^B$) is

\begin{widetext}
\bea
\frac{\frac{d\Omega_G^T(k,\eta_0)}{d\log (k)}}
{\frac{d\Omega_G^B(k,\eta_0)}{d\log (k)}} \simeq 
\frac{3}{800\pi^3}
\left(\frac{\al-2}{\al+3}\right)^2 \frac{1}{v_L^2}
\frac{\left(\frac{\Om_T(\eta_*)}{\Om_\rad(\eta_*)}\right)^2}
{\left(\frac{\Om_B(\eta_*)}{\Om_\rad}\right)^2}
\left\{  \begin{array}{l}
\frac{1}{A_1^2}\, \big(\frac{7\ga}{2}\big)^2 (L_* \mathcal{H}_*)^2
\,\,\, \text{for}~0<x<L_* \mathcal{H}_* \vspace*{0.2cm} \\
\frac{1}{A_1^2}\,x^2
\,\,\, \text{for}~L_* \mathcal{H}_*<x<1 \vspace*{0.2cm} \\
\frac{1}{A_2^2}\,x^{-(\al+3)}
\,\,\, \text{for}~1<x<(2v_L)^3 \vspace*{0.2cm} \\
\frac{4v_L^2}{A_2^2}\,x^{-(\al+11/3)}
\,\,\, \text{for}~(2v_L)^3<x<L_*/\la_* \vspace*{0.2cm} \\
0 \qquad \text{otherwise}\,.
\end{array} \right. 
\eea
\end{widetext}

\section{Conclusions}

We have analysed the generation of a gravitational
wave background by a period of turbulence in the primordial universe,
arising after a strongly first order phase transition, and by the presence 
of a primordial magnetic field. Since the early universe is an ionised plasma, 
treating pure Kolmogorov turbulence on its own is probably not
realistic. However, given the previous literature on this subject, the
analysis in Section~\ref{turbulence} is essential 
\cite{kamionkowski,kosowsky,dolgov,nicolis}. We have present the more realistic
case of MHD turbulence in Section~\ref{magnetic}, and a  
comparison of the two cases in Section~\ref{comparison}. 

The turbulent velocity field and the (causally created) magnetic field
are necessarily uncorrelated on scales larger than a given correlation scale,
which is at best equal to the cosmological horizon. We have demonstrated that
this fact, together with the property of vanishing divergence shared
by the two fields,
implies that their power spectra are blue on scales larger than the
correlation scale. This was already known in the case of causal
magnetic fields~\cite{causality}, but is new for turbulence.
A phase of turbulence in the primordial universe is
inevitably characterised by a Batchelor energy spectrum.
A Saffman spectrum is excluded, since it relies on long range correlations
that can be generated by the pressure fields~\cite{davidson}.
On small scales, we have assumed 
Kolmogorov two thirds law for the turbulent velocity correlation, 
and we have derived the turbulent
spectrum from direct integration of the correlation function.
 We have used the asymptotic behavior of the turbulence
power spectrum right up to the peak region. Therefore, the amplitude
of our result is probably not very reliable. We plan to improve this
with a numerical analysis~\cite{plan}.   

Since turbulence is an incoherent source, the analysis of
the gravitational waves generated by it is not straightforward. In
Section \ref{GWturbulence} we propose
a procedure where the time dependence of the turbulent flow is not lost
in the ensemble averaging. This time dependence is important for the
correct determination of the induced gravitational waves. Our 
method to analyse the formation of gravitational waves is based on an
heuristic model of the oscillations of the random velocity field. This
model is inspired by Richardson's energy cascade through eddies of
different sizes. 

Our work corrects an error in the dispersion relation found
in previous papers~\cite{kamionkowski,kosowsky,dolgov,nicolis} and provides the
correct dispersion relation for the free propagation of the
gravitational waves, after the turbulent source ceases to be active. 
In Ref.~\cite{kosowsky}  the sub-horizon power spectrum
of the gravitational waves, at the time when the turbulent source ceases
to be active, is given in Eq.~(44) as a function of $k$. The component
$\lag{h}_{ij}(\bk,\eta_*){h}_{ij}^*(\bq,\eta_*)\rag \propto
\delta^3(\bk-\bq)$ of the spectrum has momentum $\bk$.
However, then they set the gravitational wave frequency at time $\eta_*$
to $\omega_k$, the oscillation frequency of the turbulent source,
\bea
\omega_k &=&  (v_L/L^{1/3})k^{2/3}  ~.
\eea
This means that the gravitational wave dispersion relation at time
$\eta_*$ is the same as the dispersion relation 
of the turbulent source. However, for $\eta >\eta_*$ the gravitational
wave propagates freely, and the dispersion relation should simply be
$\omega=k$. For $\eta >\eta_*$ the gravitational wave power spectrum
changes only by redshifting with the expansion of the universe.
Therefore, in this procedure it is not clear how one can recover the
correct dispersion relation for gravity waves at 
$\eta >\eta_*$. 

The origin of this mistake is the idea that the gravity wave
frequency is determined by the frequency and not by the wave number 
of the source.  This is certainly true for spatially small sources
with infinite lifetime~\cite{weinberg}. However, in Appendix~A we show
that for a gravitational wave source which is localised in time but
not in space, the frequency of the emitted gravitational wave is
determined by the wavenumber of the source (the generic
relation between the frequency of the gravity wave and the frequency
and the wave number of its source are derived in
Ref.~\cite{brief}). This is particularly important for  
turbulence, since the turbulent eddies do not spin at the speed of
light. As a consequence, we find that the
energy power spectrum of the induced gravitational waves peaks at a frequency
corresponding to the \emph{size} of the largest turbulent 
eddies (the correlation
scale),  and not to their frequency. The spectrum 
continues to both larger and smaller wavelenghts with different
slopes. 

Usually, the nucleating bubbles of true vacuum also generate
anisotropic stresses in the fluid and therefore lead to the production
of gravity waves. The spectrum of these gravity waves from the bubbles
also peaks at frequency $L_*^{-1}$. Its amplitude has been estimated for
example in Ref.~\cite{kamionkowski}. 
 
In the case of magnetic fields, we obtain a blue power spectrum
on large scales, leaving as free parameters both the slope of the
power spectrum on small scales, and the growth of the correlation
length with time.  Assuming the existence of the invariant
quantity $\lag B^2 \rag L^5$ analougous to
Loitsyansky's integral, we obtain
 that the comoving anisotropic stress depends on time  only via the
magnetic correlation length. We have solved the evolution equation for
gravitational waves sourced by magnetic anisotropic stresses accounting for
the evolution of the magnetic correlation length. This has allowed us
to derive new limits for the amplitude of the magnetic field as a
function of the 
correlation scale, extending the results of Ref.~\cite{nucleosintesi}
to correlation scales which are smaller than the horizon.

We conclude our analysis with a comparison of the efficiency
of turbulent and magnetic sources. When imposing that the total 
energies available in the sources and their correlation lenghts be
comparable, we find that the total gravitational wave
energy densities are of the same order, but the intensity of 
gravitational waves generated by magnetic fields on super-horizon scales is
enhanced with respect to that of gravitational waves from  turbulence
by a factor $(L_*\mathcal{H}_*)^2$. We can conclude
that a coherent, relativistic, long acting source as the magnetic field
is more efficient in generating gravitational
waves on super-horizon scales, than a source which acts
only over a brief period of time like turbulence.

\section*{Acknowledgment}
We thank D. Grasso, T. Kahniashvili, A. Kosowsky, M. Maggiore, A. Nicolis and P. Wittwer for
discussions. This work is supported by the Swiss National Science Foundation.

\appendix

\section{}
\label{appendix}

In this appendix we want to clarify that, for a stochastic
gravitational wave source which is homogeneously distributed in space
and which has a time duration comparable to its characteristic 
oscillation period,
the frequency of the generated gravity wave is
connected to the wavevector ${\bf k}$ (hence to the typical size) of
the source and not to its frequency. The calculation 
is already performed for the particular case of the turbulence in section
\ref{GWturbulence}, but it is repeated here in more generality for
clarification. We neglect cosmological expansion since it is of no
relevance for our argument and consider simply a Minkowski background. 
The wave equation 
\be
(\partial_t^2-\triangle)h_{ij}(\bx,t)=8\pi G \Pi_{ij}(\bx,t)\,,
\ee
can be solved by the retarded potential
\be\label{hijspace}
h_{ij}(\bx,t)=8\pi G \int d^3x'\frac{\Pi_{ij}(\bx',t-|\bx-\bx'|)}{|\bx-\bx'|}\,.
\ee
For a long lived source confined in some small spatial region, one may consider
the usual wave-zone approximation.
In the wave zone, that is at distances much larger than the dimension of the
source, the field emitted by the source looks like an outgoing spherical wave
with wavenumber $ k= \omega $, where $\omega$ is the source
frequency $\Pi_{ij}(\bx,t)=\Pi_{ij}(\bx)e^{i\omega t}$ 
\cite{weinberg}. 

However, we are treating the case of a stochastic, statistically
homogeneous cosmological source. 
The source is not stationary, but characterized by a well defined time
interval of action. Therefore, in order to solve the above equation 
we first Fourier transform it in space
\be
(\partial_t^2+k^2)h_{ij}(\bk,t)=8\pi G \Pi_{ij}(\bk,t)\,.
\ee
The retarded Green's function of the above operator (the Fourier
transform of the Green function used in eq.~(\ref{hijspace})) is simply
\be
G_R(k,t)=\Theta(t)\frac{\sin(kt)}{k}\,,
\ee
where $\Theta(t)$ is the Heaviside function. 
If the source starts emitting at a time $t_{\rm in}$ we have
\be
h_{ij}(\bk,t)=8\pi G\int_{t_{\rm in}}^t dt' \frac{\sin(k(t-t'))}{k}
\Pi_{ij}(\bk,t')\,.
\ee
At the time $t_{\rm fin}$, the gravitational wave is given by
\bea
h_{ij}(\bk,t_{\rm fin})&=& 8\pi G\int_{t_{\rm in}}^{t_{\rm fin}} 
dt' \frac{\sin(k(t_{\rm fin}-t'))}{k}\Pi_{ij}(\bk,t') \nonumber \\
\dot{h}_{ij}(\bk,t_{\rm fin})&=& 8\pi G\int_{t_{\rm in}}^{t_{\rm fin}} 
dt' \cos(k(t_{\rm fin}-t'))\Pi_{ij}(\bk,t')\,. \nonumber
\eea
After the source stops emitting, the gravitational wave background obeys the
free wave equation $(\partial_t^2+k^2)h_{ij}(\bk,t)=0$, with solution
\be
h_{ij}(\bk,t>t_{\rm fin})=A_{ij}\frac{\sin(k(t-t_{\rm fin}))}{k}+B_{ij}\cos(k(t-t_{\rm
  fin}))\,.
\label{free}
\ee
Matching the free wave and the sourced wave at time $t_{\rm fin}$ we obtain
\bea
\hspace*{-2mm} A_{ij}(\bk)&=& 8\pi G\int_{t_{\rm in}}^{t_{\rm fin}} 
dt' \cos(k(t_{\rm fin}-t'))\Pi_{ij}(\bk,t') \nonumber \\
\hspace*{-2mm} B_{ij}(\bk)&=& 8\pi G\int_{t_{\rm in}}^{t_{\rm fin}} 
dt' \frac{\sin(k(t_{\rm fin}-t'))}{k}\Pi_{ij}(\bk,t')\,.
\eea
Regardless of the particular time dependence of the source
$\Pi_{ij}(\bk,t')$, the free gravitational field (\ref{free})
oscillates with the frequency $\om=k$ corresponding to the wavenumber
of the source. If the source is itself oscillating, but it obeys a
non-trivial dispersion relation $\omega(k)\neq k$, then the source
frequency intervenes in the amplitude of the free propagating wave. 
Only if the duration $t_{\rm fin} -t_{\rm in}$ is much larger than the
period $1/\om$, the above integrals can be approximated by
$\de(k-\om)$ and the wave inherits the frequency of the source \cite{brief}.

The same result can also be obtained from
Eq.~\eqref{hijspace}, using that $|\bx-\bx'|$ has to lie in the
interval $[t-t_{\mathrm{fin}},t-t_{\mathrm{in}}]$. 

This has been misrepresented in previous literature (see
e.g. \cite{kamionkowski,dolgov}). It is not very important for the
results obtained as long as $v_L ={\cal O}(1)$, but it is an
interesting conceptual point: A stochastic spatially homogeneous source for
gravity waves which has a lifetime comparable to its characteristic oscillation
period (as it is the case for turbulence) imprints its {\em wave
number} and not its frequency in the gravity waves it produces. On the
other hand, a stationary and isolated source 
always imprints its {\em frequency} and not its wave number.

\section{}

In this appendix we rewrite the spectra of the gravitational waves generated
by turbulence in terms of the phase transition parameters as they are
conventionally defined. We conform to the notations of Refs. 
\cite{dolgov,nicolis} which are commonly used in the literature. We denote
$u_S=v_L$ the velocity of the largest eddies, so that the total kinetic energy
of the turbulence is $\frac{\Om_T(\eta_*)}{\Om_{\rm rad}(\eta_*)}\simeq
\frac{2}{3}u_S^2$. The duration of the phase transition is
$\beta^{-1}=a_*\TT$, and $H_*$ is the physical Hubble parameter at the
phase transition. The frequency
of the gravitational waves is denoted $f=k$, and $f_S=\om_L/a_*$ is the
physical turnover frequency of the largest eddies scaled to today. The peak frequency given in
Eq.~(\ref{peakfre}) corresponds to $k_L\simeq f_S/u_S$. We call $f_{\rm
  max}=1/\la$ the maximal frequency corresponding to the dissipation
scale. Finally, the gravitational wave energy density per logarithmic
unit of frequency is denoted  $\Om_{GW}(f)$. 
We now express our results in terms of $f_S, ~u_S, ~H_*, ~\beta$ and $v_b$.

Instead of considering the full expression (\ref{long}), we find it more useful to rewrite the
approximated formulas (\ref{Om0.01}) and (\ref{Om3}). 
In the case $u_S\leq 1/2$, Eq.~(\ref{Om0.01}), we find
\begin{widetext}
\bea
\Omega_{GW}(f)\simeq  \frac{\Om_{\rm rad}}{2\pi}
\left(\frac{H_*}{\beta}\right)^2 v_b^2\,
\left\{ \begin{array}{ll}
\vspace*{0.2cm}
\frac{u_S^5}{4}\, \big(\frac{f}{f_S}\big)^3 & \text{for}~~0<f<2f_S\\
\vspace*{0.2cm}
u_S^5\, \frac{f}{f_S} & \text{for}~~2f_S<f<f_S/u_S\\
\vspace*{0.2cm}
u_S^{4/3}\, \big(\frac{f_S}{f}\big)^{8/3} & \text{for}~~f_S/u_S<f<f_{\rm max}\\
0 & \text{otherwise}\,.
\end{array} \right.
\eea
\end{widetext}

In the case $u_S\geq 1/2$, Eq.~(\ref{Om3}), we find
\begin{widetext}
\bea
\Omega_{GW}(f)\simeq  \frac{\Om_{\rm rad}}{2\pi}
\left(\frac{H_*}{\beta}\right)^2 v_b^2\,
\left\{ \begin{array}{ll}
\vspace*{0.2cm}
\frac{u_S^5}{4} \,\big(\frac{f}{f_S}\big)^3 & \text{for}~~0<f<f_S/u_S\\
\vspace*{0.2cm}
\frac{1}{4}\,\big(\frac{f_S}{f}\big)^2 & \text{for}~~f_S/u_S<f<8\,u_S^2\,f_S\\
\vspace*{0.2cm}
u_S^{4/3}\, \big(\frac{f_S}{f}\big)^{8/3} & \text{for}~~8\,u_S^2\,f_S<f<f_{\rm max}\\
0 & \text{otherwise}\,.
\end{array} \right. 
\eea
\end{widetext}

\end{document}